\documentclass[10pt,journal,compsoc]{IEEEtran}

\IEEEoverridecommandlockouts

\usepackage{url}
\usepackage{cite}
\usepackage{graphicx}
\usepackage{booktabs}
\usepackage{multirow}
\usepackage{eurosym}
\usepackage{amssymb}
\usepackage{amsmath}
\usepackage{subfigure}
\usepackage{color}
\usepackage{caption}
\usepackage{comment}

   \graphicspath{{figures/alphaVar/}{figures/meta_perc_var/}{figures/maxq_var/}{figures/substats/}{figures/fairness/}{figures/methodology/}{figures/qwait/}}
   \usepackage{subfig}

\begin{document}

\title{Metascheduling of HPC Jobs in Day-Ahead Electricity Markets \thanks{This work was supported by Department of Science and Technology (DST), India via the grant SR/S3/EECE/0095/2012.}}

\author{Prakash~Murali, Sathish~Vadhiyar,~\IEEEmembership{Senior Member,~IEEE} %
\IEEEcompsocitemizethanks{
\IEEEcompsocthanksitem Prakash Murali is doing his PhD in the Computer Science Department, Princeton University, USA. \protect\\
E-mail: pmurali@princeton.edu
\IEEEcompsocthanksitem Sathish Vadhiyar is with the Department of Computational and Data Sciences, Indian Institute of Science, India.\protect\\
Email: vss@iisc.ac.in
}}

\maketitle

\begin{abstract}
High performance grid computing is a key enabler of large scale collaborative computational science. With the promise of exascale computing, high performance grid systems are expected to incur electricity bills that grow super-linearly over time. In order to achieve cost effectiveness in these systems, it is essential for the scheduling algorithms to exploit electricity price variations, both in space and time, that are prevalent in the dynamic electricity price markets. In this paper, we present a metascheduling algorithm to optimize the placement of jobs in a compute grid which consumes electricity from the day-ahead wholesale market. We formulate the scheduling problem as a Minimum Cost Maximum Flow problem and leverage queue waiting time and electricity price predictions to accurately estimate the cost of job execution at a system. Using trace based simulation with real and synthetic workload traces, and real electricity price data sets, we demonstrate our approach on two currently operational grids, XSEDE and NorduGrid. Our experimental setup collectively constitute more than 433K processors spread across 58 compute systems in 17 geographically distributed locations. Experiments show that our approach simultaneously optimizes the total electricity cost and the average response time of the grid, without being unfair to users of the local batch systems.
\end{abstract}

\section{Introduction}

High performance grid computing involving supercomputer systems at distributed sites plays an important role in accelerating scientific advancement and facilitating multi-institutional and multi-disciplinary collaborations.  Extreme Science and Engineering Discovery Environment (XSEDE), Open Science Grid and European Grid Infrastructure are some examples of computational grid infrastructure that support science gateways to enable communities to use HPC systems.  The operational costs of these systems have become comparable to the cost of hardware acquisition, and service providers regularly budget millions of dollars annually for electricity bills \cite{Yang:2013:IDP:2503210.2503264}. With the growing demand for exascale computation, the power consumption and operational costs of these systems are expected to increase super-linearly over the years. Hence it is imperative to include power and electricity cost minimization in job scheduling decisions in high performance computational grids.

A large body of work has been developed to reduce the power consumption of data center servers, by switching off unused nodes \cite{4724317}, using voltage and frequency scaling to run servers at low power \cite{5493460}, and using renewable energy sources to reduce the carbon footprint of computation \cite{6114408}.  We consider our work of metascheduling our applications to sites to reduce time and electricity cost as complementary to these approaches.  Deregulation of the electricity power markets, creation of power trading zones, and use of renewable energy in many countries offer opportunities to purchase wholesale power under various dynamic pricing schemes.  Dynamic electricity price markets are popular and cater to large industries and manufacturing units. In such markets, scheduling algorithms can exploit spatial and temporal electricity price differentials and schedule workloads at servers which have cheap power.  

Typically, the wholesale energy market consists of a day-ahead market and a real time market. In the day-ahead market, consumers of electricity submit bids with their expected power requirements for the following day (demand), and suppliers of electricity submit bids with their expected generation and supply volumes for the coming day (supply). The trading agency which accepts these bids, sets a clearing price for each hour of the coming day, based on the supply and demand bids.  In contrast, real-time markets operate at a faster rate, and the prices can fluctuate, say every 5 minutes, based on the actual supply and demand scenario in the market. 

HPC sites and systems can also participate in such demand-response electricity programs \cite{lbnl-5763e-2012}.  With the increasing power requirements in HPC, we anticipate that in the near future, HPC system operators will consider these markets as a potential source of cheap power.  HPC sites like Argonne National Lab (ANL) are already considering using electricity according to time-of-use pricing \cite{Yang:2013:IDP:2503210.2503264}.  We use the day-ahead hourly electricity prices because the day-ahead markets are suitable for HPC workloads. The loads on these systems are predictable at a coarse level and can be used by administrators to submit accurate demand bids for procuring power supply the following day. Moreover, the prices in the day-ahead market fluctuate smoothly and can be predicted using time series forecasting techniques. These predictions can be used for intelligent scheduling decisions.

For a scheduler to estimate the total electricity price for a job execution before allocating the job to a system with hourly price variations, it is important to know the period of execution in the system. Production parallel systems in many supercomputing sites are batch systems that provide space sharing of available processors among multiple parallel applications or jobs.  Well known parallel job management frameworks including including IBM LoadLeveler\cite{ibm-load-leveler}, PBS\cite{pbs}, Platform LSF\cite{ibm-platform-lsf} and Maui scheduler\cite{maui} are used to provide job queuing and execution services for users on these supercomputers. With multiple users contending for the compute resources, a batch queue submission incurs time due to waiting in the queue before the resources necessary for its execution are allocated. The queue waiting time ranges from a few seconds to even a few days on production systems, and is dependent on the load of the system, the batch scheduling policy and the number of processors requested by the user. Thus, the queue waiting time and hence the starting time of the job on the system is not known in advance. For the execution time and the ending time of the job, we use the estimated run time (ERT) provided by the user in the job script. The ERT of the job is required for system schedulers which employ backfilling to increase system utilization, and is thus supported by many of the job management frameworks including PBS. When the user does not specify the ERT, the maximum runtime limit is assumed.  

In this work, an extension to our previous work \cite{prakash-metascheduling-hipc2015}, we have developed a metascheduling strategy that considers hourly electricity price variations in a day-ahead market and predicted response times to schedule HPC parallel jobs to geographically distributed HPC systems of a grid. Our metascheduler simultaneously minimizes electricity cost and response times by exploiting electricity price differences across states and countries to schedule jobs at systems where the cost of servicing the job is minimized while ensuring that the users do not suffer degradation in system response time. Our metascheduler uses a framework that we have developed for prediction of queue waiting times. We formulate the job scheduling problem in our metascheduler as a minimum cost maximum flow computation in a suitable flow network and use the network simplex algorithm for optimization \cite{Orlin:1996:PTP:313852.314105}. We evaluated our algorithm with trace based simulations using synthetic and real workload traces of two production grids: XSEDE \cite{xsede} and NorduGrid \cite{nordugrid}, and real electricity price data sets. Our approach can potentially save \$167K in annual electricity cost while obtaining $25\%$ reduction in average response time compared to a baseline strategy. We found that even users who do not use our metascheduler, can sometimes obtain improvements in response time when our algorithm migrates jobs away from their local systems.

To our knowledge, ours is the first work on metascheduling HPC workloads across grid systems considering actual or predicted hourly electricity prices at a predicted period of job execution.

In Section \ref{background}, we motivate and describe the problem definition. We discuss our methodology including the 
network flow formulation in Section \ref{methodology}. The experimental setup is detailed in Section \ref{expt-setup-section}. We present the results and some practical 
considerations in Section \ref{results}. We describe the related work in Section \ref{related-work} and conclude in Section \ref{conclusion}.

\section {Background}
\label{background}

Popular metaschedulers like Condor-G \cite{xsede-condor} use the concept of periodic scheduling cycles to efficiently manage job submission and dispatch decisions.  When a job is submitted by a user, the metascheduler marks the job as pending for scheduling.  During the subsequent scheduling cycle, the scheduling algorithm assigns a subset of the pending jobs for processing at a subset of the systems in the grid.  In many currently operational grids, administrators impose restrictions on the maximum number of jobs that can be submitted to a particular system in a single scheduling cycle to prevent the middleware at these systems from being flooded by job submissions \cite{xsede-condor}. We denote this maximum number as $MaxQ$.

Given $n$ geographically distributed grid systems with day-ahead hourly electricity prices and a meta scheduling portal for accepting job submissions, the metascheduling problem is to assign jobs in a scheduling cycle to systems while simultaneously minimizing the response time and electricity cost of the job executions.
  
While our metascheduler may increase the local electricity cost at a system due to job migrations from submitting to execution systems, it attempts to reduce the overall operational cost of the grid.  We also claim that the variations in workload at a particular system due to our metascheduler cannot significantly alter the day-ahead hourly electricity prices at the system's location.  This is because the  day-ahead market trading volume is typically many orders of magnitude higher than the power consumption of a single computing system. For example, the Gordon system at San Diego Supercomputer Center in California has a maximum power consumption of 358.4KW \cite{top500}, while the peak trading volume in California Independent System Operator (CAISO)  for all days in August'14 is more than 30,000MW \cite{caiso-market}. 

\section{Methodology}
\label{methodology}

We formulate the grid scheduling problem as a minimum cost maximum flow computation and use the network simplex algorithm to find the optimal flow. To compute the cost of scheduling a job on a system, we require predictions of the response time of the job at the system and the electricity cost required to execute the job. We describe our approach for prediction of response time in Section \ref{resp-pred-section}, and prediction of electricity price in Section \ref{price-pred-section}. In Section \ref{mcmf-section}, we define the cost function and the flow network used in our approach.

\subsection{Response time prediction}
\label{resp-pred-section}
In batch queue systems, similar jobs which arrive during similar system queue and processor states, experience similar queue waiting times. We have developed an adaptive algorithm for prediction of queue waiting times on a parallel system based on spatial clustering of the history of job submissions at the system \cite{prakash-qespera-cpe2016}. To obtain the prediction for a job $J$ on a system $S$, $J$ is represented as a point in a feature space using the job characteristics (request size, estimated run time) specified by the user, the queue state at the system at the current time (sums of request sizes of queued jobs, estimated run times of queued jobs, elapsed waiting time of queued jobs) and the state of the compute nodes at the current time (number of occupied nodes, total elapsed running time of the jobs, total estimated run times of the jobs). We compute the Manhattan distance of each history job with the target job, and consider history jobs with small distance values as being similar to the target job. Then, we use Density Based Spatial Clustering of Applications with Noise (DBSCAN) to find clusters of similar jobs. DBSCAN also allows us to eliminate outliers among the history jobs. If we find clusters which are very similar to the target job, i.e., clusters with low average distance,  we use the weighted average of waiting times of jobs in the cluster as the prediction for the job, $J$.  If we do not find clusters which are very similar to the job, the job features of the history jobs and the queue waiting times experienced by these jobs are used to train a ridge regression model.  Using an iterative least squares minimization, ridge regression obtains a linear model which is robust to the ill-conditioning present in our feature matrix. The features of the target job are supplied as input to this model to obtain the predicted queue waiting time.

To find the response time of a job on a target system, we invoke our queue waiting time predictor to find the predicted start time of the job, $t_s$. Then, we use the estimated run time (ERT) supplied by the user to predict the end time of the job. While the user estimates are known to be inaccurate \cite{Tsafrir:2007:BUS:1263127.1263243}, the estimates serve as strict upper bounds on the runtimes since job schedulers used in HPC systems terminate a job when its runtime exceeds the user estimated runtime. In this work, we use these estimates to demonstrate the benefits that can be obtained for the grid systems from participation in dynamic electricity markets. We show in our experiments that using these estimates results in improved scheduling decisions over a strategy that does not use predictions, but only considers the loads at the time of the submission. We expect that using improved runtime prediction strategies can lead to additional benefits.
  
Since the ERT supplied by the user is relevant only for the submission system, we use a scaling factor to adjust the ERT for the target system. This scaling factor is computed by taking the ratio of the performance per core (in GFlops) of the target system and the submission system. For a job which is submitted at a system $S_i$, for which we require an estimate of the runtime at system $S_j$, we obtain the performance per core of both systems, and scale the ERT of the job as $ERT_{S_j} = ERT_{S_i}\times ppc_{S_i}/ppc_{S_j}$ where $ERT_{S_i}$ is the estimated run time of the job provided by the user on system $S_i$, $ppc_{S_i}$ is the performance per core of system $S_i$. The predicted end time of the job on the system $S_j$ is $t_e = t_s + ERT_{S_j}$. We describe our approach for estimating the power per core of a system in Section \ref{expt-setup-section}.

Migration of jobs from submission to execution sites involves transfer of data and executables. In practice, the data size parameter can be given as input by the user, and the cost of data movement can be computed using the data size and the properties of the link (latency and bandwidth) between the submission and the execution sites.  We can also predict the data transfer delays using regression based techniques \cite{Vazhkudai:2003:URT:1080646.1080678} and include them in our calculations of response times. We can also employ just-in-time data transfer techniques \cite{monti-jitstagging-petaworkshop2008} to overlap the data transfer time and queue waiting time, and mitigate the impact of file transfers on job response time.  However, in this study, we do not consider data transfer times because our workloads do not include the necessary information about job file transfers and network state, and the current workload models \cite{pwa,Lublin:2003:WPS:963932.963939,Tsafrir:2005:MUR:2146214.2146215} for synthetic logs do not generate data sizes. We assume that the executable binaries and data needed for a job are set up at multiple systems prior to the job submission and hence the cost of job migration between the systems is negligible.

\subsection{Electricity price prediction}
\label{price-pred-section}
To obtain the electricity prices during the job's execution period at a target system, we find the predicted start and end time of the job using our response time predictor. Given the predicted start and end times of the job on the system, we check whether the job's predicted execution duration is within the end of the day (midnight). In this case, the corresponding electricity prices during the execution period in the day-ahead electricity market are known. When the execution period does not fully lie within the hours of the current day, i.e., $t_e$ is after midnight on the submission day, we predict the prices for the duration that lies beyond  midnight. We use a Seasonal Autoregressive Integrated Moving Average (SARIMA) model to  model the electricity prices fluctuations in the day-ahead market. SARIMA models are commonly used to obtain forecasts for time series data which exhibit seasonal trends across days and months. Since we observed that the prices in the day-ahead market have high lag-24 autocorrelation, we use the SARIMA model with a seasonal period of 24 hours. The various parameters required for the SARIMA model were tuned using a training set of the electricity price data. We used unit order terms for the autoregressive and moving average seasonal and non-seasonal components of the model for our experiments.

\subsection{MCMF: Minimum Cost Maximum Flow}
\label{mcmf-describe}

Minimum cost maximum flow (MCMF) is a fundamental network flow model which aims to maximize the amount of shipment of a single commodity through a network while minimizing the cost of the shipment.  MCMF can be solved using a number of approaches including cycle canceling, linear programming and network simplex algorithms.  We first define the minimum cost flow (MCF) problem and use it to define the minimum cost maximum flow (MCMF) problem. The MCF problem is defined as follows. Let $G(V,E)$ be a flow network with source vertex $s \in V$ and sink vertex $t \in V$.  Each edge $(u,v) \in E$ has capacity $c(u,v) > 0$, flow $f(u,v)$ and cost $p(u,v)$.  In other words, the capacity, flow and cost are mappings from $E \to \mathbb{R}^+$.  The capacity of the edge denotes the maximum flow possible along the edge and the cost denotes the price of unit flow along an edge. The flow network, cost and capacity mappings are input for the problem and the flow mapping, $f$, is the output.

Given some required flow value $d$ from $s$ to $t$, the problem $MCF(G, c, p, d)$ is
\begin{equation}
\small
 \textbf{min} \sum_{(u,v) \in E}{p(u,v)\cdot f(u,v)}
\end{equation}
subject to the following flow constraints:
\begin{subequations}
\small
\begin{align}
& \textbf{Capacity}:f(u,v) \le c(u,v), \\
& \textbf{Skew symmetry}:f(u,v) = -f(v,u), \\
\begin{split}
& \textbf{Flow conservation}:  \\
& \sum_{u:(u, v) \in E} f(u,v) = \sum_{u:(v, u) \in E} f(v,u) \; \forall u \in V \setminus \{s, t\}\\
\end{split} \\
& \textbf{Required flow}:\sum_{(s, v) \in E} f(s,v)  =  \sum_{(v, t) \in E} f(v,t)  = d.
\end{align}
\end{subequations}

The capacity constraints ensure that no edge is the network can transport more flow than its capacity.  The skew symmetry condition allows algorithms to reduce the flow on an edge by introducing non-zero flow in the reverse direction.  The flow conservation condition states that all flow which enters a vertex through its incoming edges, leaves it through the outgoing edges.  Essentially, it means that no vertex is allowed to store or buffer the flow. In addition to these three constraints, the required flow constraint is necessary to obtain a non-trivial solution for the minimization problem.  A flow function $f$ which satisfies these constraints is termed \textit{feasible}.  A feasible solution to the MCF problem is dependent on the value of $d$. Given a certain value of $d$ for which a feasible minimum flow exists, we can obtain a solution for a smaller required flow value $d'$, by reducing the flow on the edges in the network.  But, it may not be possible to find a minimum cost flow for a higher required flow value because of the capacity constraints on the edges. Hence, the Minimum Cost Maximum Flow (MCMF) problem is to find the maximum value of $d$ which can produce a feasible flow in the corresponding MCF problem. Formally, $MCMF(G, c, p)$ is The Minimum Cost Maximum Flow (MCMF) problem is to find the maximum value of $d$ which can produce a feasible flow in the corresponding MCF problem. Formally, $MCMF(G, c, p)$ is
\begin{equation}
\small
 \textbf{max} \left\{ d \in \mathbb{R}^+: MCF(G, c, p, d) \; \text{has a feasible solution.}\right\}
\end{equation}

Minimum cost flow problem can be solved using linear programming because the objective function and the constraints are linear.  Given integer capacities and costs, a solver for MCF can be used to compute the maximum feasible value of $d$ by using a binary search on the set of integers up to the total outgoing capacity of the source vertex. The network simplex algorithm \cite{Orlin:1996:PTP:313852.314105} relies on the observation that the minimum cost flow problem has at least one optimal spanning tree solution i.e., the set of edges with non zero flow form a spanning tree for the flow network. In each iteration, the algorithm pivots from one spanning tree solution to the next by replacing a tree-arc with a non tree-arc, in a manner that resembles the simplex algorithm for linear programming.  The network simplex algorithm runs in $O(\text{min}(n^{2}m \log nC, n^{2}m^{2} \log n)))$, where $n$ is the number of nodes, $m$ is the number arcs and $C$ is the maximum cost on any arc  \cite{Orlin:1996:PTP:313852.314105}. Since the optimized implementations of network simplex algorithm are usually very fast in practice, we adopt it for our research.

\subsubsection{Metascheduling using MCMF}
\label{mcmf-section}
   
To schedule a set of jobs to a set of systems, we represent the jobs and systems as nodes in a flow network.  We consider a system to be \emph{compatible} for a job, if the total cores in the system is more than the request size of the job and the maximum wall time permitted in the system is more than the user estimated run time of the job.  For each job, we add an arc of unit capacity from the job to each compatible system. A flow of unit value along an arc from $J$ to $S$ represents scheduling $J$ on $S$. We assign the cost of each job-system arc as a weighted linear combination of the predicted response time of the job and the electricity price required to execute the job on the system. 

To compute the cost of assigning a job $J$ to a system $S$, we predict the start time and end time of $J$ on $S$ as $t_s$ and $t_e$, respectively.  Assuming that the job is submitted at time $t$, the response time of the job is $T_J = t_e - t$.  In each scheduling cycle, the metascheduler polls each system in the grid to obtain its current queue and processor state in order to invoke the response time predictor for each job on each compatible system.  Using these predictions for each job on each system, we find the maximum and minimum predicted response times in this scheduling cycle as $T_{max}$ and $T_{min}$. The cost of scheduling the job $J$ at the system $S$, in terms of response time, is defined as 
\begin{equation}
\small
C_T(J, S) = (T_J - T_{min})/(T_{max} - T_{min}). 
\end{equation}
We model the electricity prices at the location of the system $S$ to obtain a function $\hat{\phi}_S(t)$ which gives the predicted electricity price during the time $t$.  The cost of scheduling the job at this system, in terms of electricity price is defined as
\begin{equation}
\small
C_E(J, S) = \frac{\sum_{t=t_s}^{t_e} P_{J,S}\times\Delta\times\hat{\phi}_S(t) - E_{min}}{E_{max} - E_{min}} 
\end{equation}
where $P_{J,S}$ denotes the power consumption of $J$ on $S$, $\Delta$ denotes the period of the day-ahead electricity market, and $E_{max}$ and $E_{min}$ denote the maximum and minimum predicted electricity cost observed in the current scheduling cycle. Since we use prices from the day-ahead hourly market, $\Delta=$ 1 hour in all our experiments. We describe our approach to calculate $P_{J,S}$ in Section \ref{expt-setup-section}. 

We define the cost of scheduling $J$ on $S$ as
\begin{equation}
\small
C(J,S) = w_t\times C_T(J,S) + (100-w_t)\times C_E(J,S)
\label{weightedComb}
\end{equation}
where $w_t$ is the relevance of the response time in the cost function.  $C_T(J,S)$ and $C_E(J,S)$ are normalized by the corresponding minimum and maximum values to unit-less quantities so that $w_t$ can be used for weightage of the two terms irrespective of their absolute value. Our formulation shown in Equation \ref{weightedComb} is based on the nadir-utopia normalization method by Kim and Weck \cite{kim-adaptiveweighted-smo2006}. In every scheduling cycle, the objective function is re-normalized to adapt it to the current predictions of electricity price and queue waiting time.
 
\begin{figure}
\centering
\includegraphics[scale=0.75]{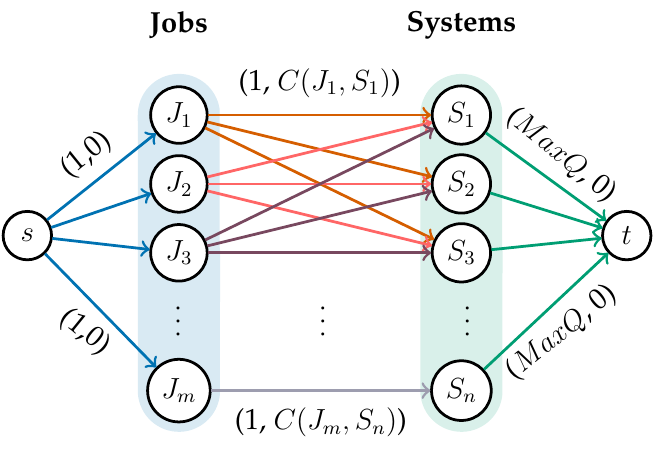}
\caption{Flow network used for scheduling. The edges are labelled as (edge capacity, cost of unit flow).}
\label{maxflowfigure}
\end{figure}

We connect an arc of unit capacity from the source node $s$ to each job and an arc of capacity $MaxQ$ from each system to the sink node $t$. The costs of these edges are set to 0.  For a set of $m$ jobs and $n$ systems, we illustrate this network in Figure \ref{maxflowfigure} where each arc is labeled with two parameters, namely, the capacity of the edge and the cost of unit flow through the edge.  We scale the costs of the network edges by multiplying with a large constant (100), and round off the values to integers. In such a network, the Integrality Theorem of maximum flow networks \cite{Kleinberg:2005:AD:1051910} guarantees that the maximum flow is integral and each unit capacity edge in our network has a flow value of either 0 or 1. We compute a maximum flow of minimum cost in this network using the network simplex algorithm available in the Python package, {\em NetworkX}.  After computing the minimum cost flow, we inspect the job-system arcs and select the arcs which have non-zero flow.  For each arc from $J$ to $S$ which has non zero flow, we schedule the job $J$ on system $S$.  By the flow conservation principles, we are guaranteed that a) not more than one system is selected for a job and b) no system receives more than $MaxQ$ jobs during one scheduling cycle.

\subsubsection{Comparison with Stable Matching Algorithm}
\label{simple-comp}

In this section, we describe an algorithm for metascheduling based on stable matching and discuss why such an approach is not suitable for our metascheduling problem. 

We consider the scheduling algorithm used by HTCondor \cite{hcond} which negotiates job and machine requirements to arrive at an optimal schedule. In this approach, users submit their job requirements and daemons running at the grid systems submit details about the available resources to the metascheduler.  Users and machines are allowed to provide priority calculation methods to rank machines and jobs, respectively.  In each scheduling cycle, the metascheduler uses this information to create a preference list for each job and machine.  For example, each job can rank the machines in the decreasing order of their processing speed.  Using these preference lists, the metascheduler conducts a series of proposals where jobs propose their allocation on a machine, and the machine either accepts or denies the request.  The proposals follow two simple rules:
 \begin{itemize}
  \item no job proposes to a particular machine more than once,
  \item a machine which is free accepts any incoming allocation proposal.
 \end{itemize}
When a job proposes to a free machine, the machine accepts the proposal and the job is marked as tentatively scheduled on the machine, which is then marked as busy. If a job proposes to a busy machine, the machine can reject or accept the proposal depending on the relative priorities of the incoming and the currently scheduled job.  If the machine accepts the incoming job in favor of the currently scheduled job, the incoming job is tentatively scheduled, and the current job is tentatively marked as unscheduled.  Since each job proposes to a machine only once, the algorithm terminates in $O(n^2)$ iterations for $n$ jobs and $n$ machines. This algorithm produces a \textit{stable} matching. A matching is termed stable, if there is no pair of jobs $J_1$ and $J_2$ matched to $M_1$ and $M_2$ respectively, such that job $J_2$ prefers $M_1$ over $M_2$ and system $M_1$ prefers $J_2$ over $J_1$.  In other words, there is no job $J_2$ and machine $M_1$ which prefer each other, but are matched to another choice.

HTCondor does not use any power or electricity price optimization criteria while negotiating a schedule. However, it can be extended to include these criteria. Jobs can rank machines in the increasing order of predicted response time and machines can rank jobs in the increasing order of estimated electricity cost for the job execution. Then, the algorithm can negotiate a schedule where jobs prefer systems where the response time is less and machines prefer jobs which will incur less electricity bills.  However, this approach has multiple drawbacks.  One  drawback of this approach is that the schedule that is generated can be either job-optimal or machine-optimal, depending on if the jobs propose to machines or vice-versa.  In stable matching literature, it is well known that if the jobs propose to machines, the schedule will match each job to its best possible machine and each machine to its worst possible job \cite{gs1962}. This solution is job-optimal. When the machines propose to the jobs, the matching obtained will by machine-optimal where each machines are matched to the best possible job.  Another drawback is that if each job has a different machine as its top preference, then the algorithm will not consider the machine preferences at all. Each job will be assigned to its top preference machine and the algorithm will be incapable of optimizing electricity prices.  

A third issue with this approach is that ranking of jobs/machines loses information that can be vital for obtaining good schedules. For example, consider a job $J_1$ which has response times (1s, 10s and 20s) on three machines $(M_1, M_2, M_3)$  and a job $J_2$ which has response times (50s, 1 day, 2 days) on the same three machines.  In supercomputer systems, it is known that response times less than 20 minutes encourage the user to continue his job submission session \cite{feitelson09}. Both jobs will rank the machines in the same order and we will lose the information that from the perspective of user productivity, assigning $J_2$ to its second preference is worse than assigning $J_1$ to its second preference. Hence, it is necessary to use the exact values of predicted response time and electricity cost. In our proposed approach,  the objective function includes the exact values of both response time and electricity price for minimization. Since our approach is a global minimization in the flow network, our approach can simultaneously optimize the metrics of interest for both jobs and machines, and avoids the potential drawbacks of the HTCondor approach.

\section{Experimental Setup}
\label{expt-setup-section}

We performed trace based simulation of real and synthetic grid workloads using real electricity price data sets and power consumption profiles of compute systems to test the effectiveness of our approach\footnote{Our simulator, metascheduler, predictors and data sets are available in \url{https://github.com/MARS-CDS-IISc/mcmf-metascheduler-predictors}.}. In this section we explain each component of the experimental setup.

\subsection {Workload and Scheduler Simulator}
We conduct simulations using grid traces which are in Standard Workload Format (SWF) \cite{pwa} or Grid Workload Format (GWF) \cite {gwa}. Each line in the SWF/GWF trace denotes a job and records the arrival time, run time, number of cores, user estimated runtime and other job parameters. GWF traces also record the system where the job was originally submitted. While using SWF traces during synthetic trace generation, we appended each line in the trace with the submission system.  To conduct trace driven simulations, we used an extended version of the Python Scheduler Simulator (pyss) developed by the Parallel Systems Lab in Hebrew University \cite{pyss}.  pyss accepts a workload trace, system size and scheduling algorithm as inputs and replays the job arrival events, start and end of job execution events to simulate the state of the system with the input workload. Since pyss simulates only one system, we extended it to support multiple systems. We implemented a metascheduler class that acts as a common interface between the job submissions and the various execution systems.  We configured pyss to use the EASY backfilling algorithm \cite{Feitelson:2004:PJS:2128864.2128865} to schedule jobs at the individual systems.  

\subsection {Grid systems}
We simulate two currently operational grids which collectively span 58 individual compute systems, 17 countries and states, 10 electricity transmission operators, 7 time zones and more than 250k job submissions. For each system, we obtained the number of cores and maximum wall time of a job from publicly advertised system information.

\subsubsection{XSEDE}
The Extreme Science and Engineering Discovery Environment (XSEDE) project is a large scale compute grid which connects many universities and research centers in the US. For high performance computing, XSEDE connects eight supercomputing systems situated across different states in the US. For our simulation experiments, we used eight CPU-only systems of XSEDE and its previous incarnation, TeraGrid.  The XSEDE system configuration, shown in Table \ref{XSEDE-Site-Table} was obtained from \cite {xsede} and \cite{teragrid-sites}.  Each individual XSEDE system uses the Portable Batch System (PBS) or Sun Grid Engine (SGE)  for job management, and grid submissions are processed through Condor-G metascheduler \cite{xsede-condor}.  Hart \cite{Hart:2011:MTW:2076556.2076563} reports that short jobs which run for less than 30 minutes constitute less than $5\%$ of the total core hours delivered by XSEDE.  These jobs include debugging and test runs and are not representative of the production usage.  We do not consider these jobs for simulation and model only the jobs in the production workload.

\begin{table}
\scriptsize
\centering
\caption{\small The XSEDE Grid}
\begin{tabular}{@{}llp{0.4in}p{0.6in}p{0.6in}@{}}
\toprule
System     & Location & Cores & Power (watts/core) & Performance (Gflops/core) \\ \midrule
Blacklight & Pittsburgh                         & 4096                                                                                                           & 87.89                                                                            & 9.00                                                                                    \\
Darter     & Tennessee & 11968                                                                                                           & 30.58                                                                            & 20.79                                                                                   \\
Gordon     & San Diego                           & 16160                                                                                                          & 22.17                                                                            & 21.10                                                                                   \\
Trestles   & San Diego & 10368                                                                                                           & 42.66                                                                            & 9.64                                                                                    \\
Mason      & Indiana & 576                                                                                                             & 39.95                                                                            & 7.43                                                                                    \\
Lonestar   & Texas & 22656                                                                                                           & 15.83                                                                            & 13.32                                                                                   \\
Queenbee   & Louisiana & 5440                                                                                                            & 16.25                                                                            & 9.37                                                                                    \\
Steele     & Purdue & 4992                                                                                                            & 83.75                                                                            & 13.33                                                                                   \\ \bottomrule
\label{XSEDE-Site-Table}
\end{tabular}
\end{table}

\subsubsection{NorduGrid}
NorduGrid is a very large grid with 80 systems spread across 12 countries with a majority of the systems located in the Nordic countries. We simulated 50 selected systems of NorduGrid which constitute over $90\%$ of the total CPU cores available in the grid. In particular, we ignore systems with very small number of cores and systems where we could not obtain electricity price or CPU architecture information which is required for our simulations. The grid configuration was obtained from \cite{nordugrid}.

 \subsection {Workload}
For simulating the jobs at a system, we used a synthetic workload for XSEDE and real workload traces for NorduGrid. 

\subsubsection{XSEDE}

For the six existing systems [systems 1-6 in Table \ref{XSEDE-Site-Table}] , we obtained the system size and maximum walltime of a job at each system from their respective webpages.  For the remaining two systems, we obtained the required information from \url{www.teragridforum.org/mediawiki/images/5/5f/RPQueue_Info.xls}. The names of the systems and the system configuration used at each system are shown in Table \ref{XSEDE-Site-Table}.

We generated synthetic workload traces for each system using the workload models available in Parallel Workloads Archive \cite{pwa}. For generating the job arrival time, request size and run time, we use the model proposed by Lublin and Feitelson\cite{Lublin:2003:WPS:963932.963939}.  This is a widely used model which employs a hyper-Gamma distribution for generating job runtimes, arrival times and requests sizes.  To generate the user estimates of runtime, we used the model proposed by Tsafrir et.al. \cite{Tsafrir:2005:MUR:2146214.2146215}.  This work observes that the number of common user runtime estimates at an HPC system is usually less than 20. Their model generates practically usable estimates which mimic the modality seen in real workloads.  The model requires two inputs: the maximum value of the user estimate at a system and the number of jobs for which the estimates are to be generated. For both XSEDE and NorduGrid, we obtained the maximum values of user estimates for each system from publicly advertised system information.

In \cite{Hart:2011:MTW:2076556.2076563}, Hart provides various summary statistics about the run times, job sizes and inter-arrival times of the production jobs in XSEDE/TeraGrid.  For generating the workload, we generate traces for each system in the grid and merge them into a single workload with annotations for denoting the original submission system. We manually adjusted the parameters of Lublin and Tsafrir models to match the aggregate statistics of the synthetic workload with the reported XSEDE/TeraGrid statistics. Statistics of our synthetic workload match the characteristics reported by Hart. The average job runtime in our workload is 8.8 hours while the actual average runtime in TeraGrid is 9 hours. The average number of job arrivals at a system per hour is 3.22 in our workload, while the actual value is 3.27.

For Mason and Steele which allow long running jobs up to many weeks, we limited the maximum runtime of a job to three days because the electricity prices datasets are of limited duration.

\subsubsection{NorduGrid}
In NorduGrid, we used a real workload available in Grid Workloads Archive \cite{gwa}.  The archive records each job's submission system, submission time, requested processors and runtime. We used Tsafrir model with the maximum observed runtime as input parameter to assign user estimated runtimes for each job. 

For queue waiting time and electricity price predictions, we used a subset of jobs and electricity price data as training information. Queue waiting time is predicted for a job at a system by considering the previous 2000 job submissions at the system as the training input.  For predicting the electricity prices, we use the prices of the previous three days as training input for the SARIMA model. In Table \ref{Grid-Experiment-Configuration}, we show our simulation configuration including the number of jobs and the duration that is simulated for each grid.

\begin{table}[h]
\scriptsize
\centering
\caption{\small Simulation Configuration}
\begin{tabular}{@{}lllll@{}}
\toprule
          &       &        & \multicolumn{2}{c}{Test configuration} \\ \cmidrule{4-5}
Grid      & Systems & Cores  & Jobs                & Days             \\ \midrule
XSEDE     & 8     & 76256  & 10000               & 15               \\
NorduGrid & 50    & 356856 & 126344              & 90               \\ \bottomrule
\label{Grid-Experiment-Configuration}
\end{tabular}
\end{table}

\subsection{Variable electricity prices}
For different systems, we obtained the hourly electricity prices in the day-ahead market from the electricity operator in the respective power market. For regions without variable electricity pricing, corresponding to one system in XSEDE and 24 systems in NorduGrid, we used the applicable fixed industrial electricity price.  For systems in XSEDE, we used historical market prices for June 2014 available from the regional energy operators of the Federal Energy Regulatory Commission \cite{ferc}.  In NorduGrid, we obtained the market prices for Denmark, Sweden, Norway, Finland, Latvia and Lithuania from Nord Pool Spot \cite{nordpoolspot} and for Slovenia from BSP SouthPool\cite{bspsouthpool}. For systems in United Kingdom, Ukraine, and Switzerland, we used the applicable fixed industrial prices. Overall, in NorduGrid, our electricity price data set spans three months from January-March 2014 and includes variable electricity prices for 26 systems.

\subsection{Job power consumption and execution characteristics}
We require estimates of job power consumption and runtime at each system to quantify the impacts of job migration on metrics relevant for users and system administrators. When a job is migrated away from the submission system, the runtime and power consumption of the job can change e.g if a job is migrated to a faster system, the run time is expected to reduce. To account for these differences we require estimates for power consumption and runtime of a job at every compatible system.

\subsubsection{Job power consumption}
\label{ppc-method}
To estimate the power consumption of a job at a system, we assume that the job has the same power consumption characteristics as High Performance Linpack (HPL).  The work by Kamil et.al. \cite{4536223} experimentally demonstrates that the HPL power consumption can be used to closely approximate the power consumption of production scientific applications.  For each system in XSEDE, we obtained the peak power consumption from Top500 and Green500 datasets, computed the HPL power consumption per core, and scaled it by the number of cores used by a job to find the power consumption of a job.  Thus, if a job requires $n$ cores on a machine which has a total of $N$ cores and advertised HPL power is $P_{HPL}$, the job power consumption is considered as $P_J = n\times(P_{HPL}/N)$.  Table \ref{XSEDE-Site-Table} shows the values of HPL power consumption per core for each system.  For NorduGrid, we were unable to obtain HPL benchmark data on each system. Hence, we resorted to white papers published by the chip manufacturers to obtain the power consumption per core.  Similar to XSEDE, we scaled these numbers with the requested number of cores to find the power consumption of each job.

\subsubsection{Runtime scaling}

We assume that the applications in our workload have similar scalability characteristics as HPL.  For XSEDE, we obtained the HPL peak performance (Rmax in TFlops) of a system using Top500 data and normalized it by the number of cores in the system to find the performance per core. For a pair of systems $S_i$ and $S_j$, we compute the scaling factor $scale_{ij}$ as the ratio of the performance per core for $S_i$ and $S_j$. When a job is migrated from $S_i$ to $S_j$, we adjust the job's estimated runtime as $r_j = r_i \times scale_{ij}$ where $r_i$ is the estimated runtime of the job in system $S_i$. We computed such scaling factors for every pair of systems using Top500 data.  For NorduGrid, we obtained the theoretical peak performance of a core in the system (in GFlops) from architecture white papers. For example, for the Intel E5-2600 series of processors, we obtained the peak performance data from \cite{intelxeon}. As in XSEDE, we used these peak performance indices to compute the scaling factors across systems.

For HPL, it is reasonable to scale runtime across systems only using the number of cores, and not use other factors including memory bandwidth and communication performance. This is because the runtime of HPL is primarily dominated by the computation time $(O(N^3)$ and less by the communication times $(O(N^2)$ \cite{pfeiffer-modelingandpredicting-ipdps2008}. The computation time scales linearly with the number of processor cores. This is also confirmed by a recent study, where the memory bandwidth was found to have no impact on HPL performance, and the impact due to communication bandwidth and latency were found to be negligibly small \cite{hplinpack}. This is further confirmed by the $R_{max}$ HPL performance of large-scale systems considered in our study, where $R_{max}$ of the Top500 systems are typically about 90\% of $R_{peak}$, which is found solely using the number of cores.  Modern day applications for large-scale systems acheive or are developed to achieve linear scalability, similar to HPL. Hence, we use only the number of cores to scale the runtime to a different system.

\subsection {Evaluation metrics}
To evaluate the benefits of our approach, we employ a number of metrics as described below.
 
\begin{itemize}

\item \textbf{Average response time}

\item \textbf{Total electricity cost}:
For a job $J_i$ which executes on system $S_j$, the electricity cost is computed as:
\begin{equation}
e(J_i, S_j) =  \sum_{t={T_{W_{ij}}}}^{{T_{W_{ij}} + T_{R_{ij}}}} P_{ij}\times\Delta\times\phi_{S_j}(t)
\end{equation}

where $P_{ij}$ is the power consumption of $J_i$ on $S_j$ in watts, $\Delta$ denotes the period of the day-ahead electricity market, $T_{W_{ij}}$ and $T_{R_{ij}}$ are the waiting time and running incurred for $J_i$ on $S_j$ in hours, and $\phi_{S_j}(t)$ is the hourly price variation function for $S_j$ expressed in currency per watts. For the day-ahead hourly market, $\Delta$ is set to one hour.

\item \textbf{Bounded slowdown}:
Slowdown is computed by normalizing the response time of a job by the running time.  Since slowdown is sensitive to jobs with very small runtime, bounded slowdown thresholds the runtime using a lower bound \cite{Feitelson:2001:MPJ:646382.689681}. For a job with waiting time $T_W$ and running time $T_R$ in seconds, bounded slowdown is defined as
\begin{equation}
BS = max\left\{ \frac{T_W + T_R}{max({T_R, 10})}, 1\right\}
\end{equation}

\item \textbf{System utilization}:
Utilization at a particular system is computed by dividing the sum of the CPU hours of jobs scheduled at the system by the product of the makespan and total processors available in the system. Thus, utilization aims to measure the fraction of the system core hours which delivered useful work.

\item \textbf{System instantaneous load}: Instantaneous load is defined as the sum of the CPU hours of both the running and queued jobs divided by the total processors available in the system at a particular instant. 

\item \textbf{Fairness to System}:
The annual reports published by various supercomputing service providers which are part of XSEDE, show that, the response times of jobs processed at the system, and the number of core hours delivered to specific project allocations and users, are considered important metrics for quantifying the quality of service of each provider.
Hence, it is important for service providers to ensure that their participation in the grid does not adversely affect the users of the local batch system.  A system or site's participation in grid should not affect the quality of service provided to the jobs that are submitted to the system. Specifically, a high speed system after joining the grid may become highly loaded due to migration of jobs submitted at low speed systems. To evaluate the quality of service, we compute the speedup obtained due to the use of our metascheduling algorithm, compared to the baseline. Specifically, for a job $J$, which is submitted at system $S_i$, which has response time $R^{local}_J$ when metascheduling is not used and $R^{grid}_J$ in the presence of metascheduling, we compute the quality of service offered to the job as:

\begin{equation}
qosScore(S_i, J) = \frac{R^{local}_{J}}{R^{grid}_{J}}
\end{equation}

We then compute the fairness score for a system as the geometric mean of the QoS scores of all the jobs submitted at the system. If a system has high fairness score, it indicates that the users of the system are benefitted by the system's participation in the grid.
\end{itemize}

\section{Results and Discussion}
\label{results}
In this section, we present various results on our metascheduling approach including reduction in response time, savings in cost, and overall statistics.  We refer to our approach based on the Minimum Cost Maximum Flow algorithm as {\em MCMF}.  During our experiments, we observed that our Python implementation running on an Intel Core i7 3.4Ghz processor with 16GB RAM takes $8.4$ seconds on average for computing the scheduling cost and constructing the flow network, and $16.3$ seconds on average for computing the minimum cost flow and the subsequent job submissions to individual systems. We compare our strategy with a baseline strategy {\em BS}, in which the jobs are executed at the submission system.
 
Our strategy is primarily different from existing efforts \cite{Yang:2013:IDP:2503210.2503264, 5461933} in terms of using waiting time predictions to estimate the benefits in response time and electricity cost for the execution period of a job, and in terms of using the hourly electricity prices during the execution to estimate the total cost. Hence we compare our approach with two strategies, the first strategy called {\em INST} which does not consider predictions but makes decisions based on instantaneous loads of the systems at the time of the job submissions, and the second strategy called {\em TWOPRICE} which does not consider hourly prices but considers only two prices per day, namely, on-peak and off-peak. The INST strategy which assumes immediate execution start of a job is implemented by feeding the waiting times as zeros to our MCMF strategy. The TWOPRICE strategy is implemented by considering the on-peak hours as 12pm to midnight and calculating the off-peak and on-peak prices as the 10th and 90th percentile of the day-ahead market prices for the simulation period. Note that the INST and TWOPRICE strategies are grid scheduling strategies since they allow sharing and migration of jobs across the grid systems.  

\subsection{Prediction Accuracy}

Our MCMF metascheduler uses predictions for three parameters, namely, queue waiting times using our wait-time predictor \cite{prakash-qespera-cpe2016}, response times using user estimated runtime (ERT), and electricity prices using SARIMA model. In this section, we evaluate the prediction accuracies for the queue waiting time and electricity price predictions, and the usefulness of these predictions for metascheduling. The user estimated runtimes (ERTs) are generally known to have gross over-approximations and hence have large prediction errors \cite{Tsafrir:2007:BUS:1263127.1263243}. Section \ref{sensitivity_predictions} shows the sensitivity of our metascheduler to the errors in these predictions.

\subsubsection{Queue Waiting Time Prediction}

We evaluated our queue waiting time prediction framework using production supercomputer workload traces with varying site and job characteristics, including two Top500 systems, obtained from Parallel Workloads Archive \cite{pwa}. The detailed results and analyses are contained in our previous work \cite{prakash-qespera-cpe2016}. In summary, our predictions results in up to 22\% reduction in the average absolute error and up to 56\% reduction in the percentage prediction errors over existing strategies including QBETS \cite{Nurmi:2007:QQB:1254882.1254939} and IBL \cite{10.1109/CCGRID.2006.57} across workloads. Our prediction system also gave accurate predictions for most of the jobs. For example, for the workload of ANL's Intrepid system our predictor gave highly accurate predictions with less than 15 minutes absolute error for more than 70\% of the jobs. Our predictor is currently deployed on an 800 core system in our home department, delivering queue waiting time predictions to users with less than 30\% error.

\begin{figure}
\centering
  \includegraphics[scale=0.35]{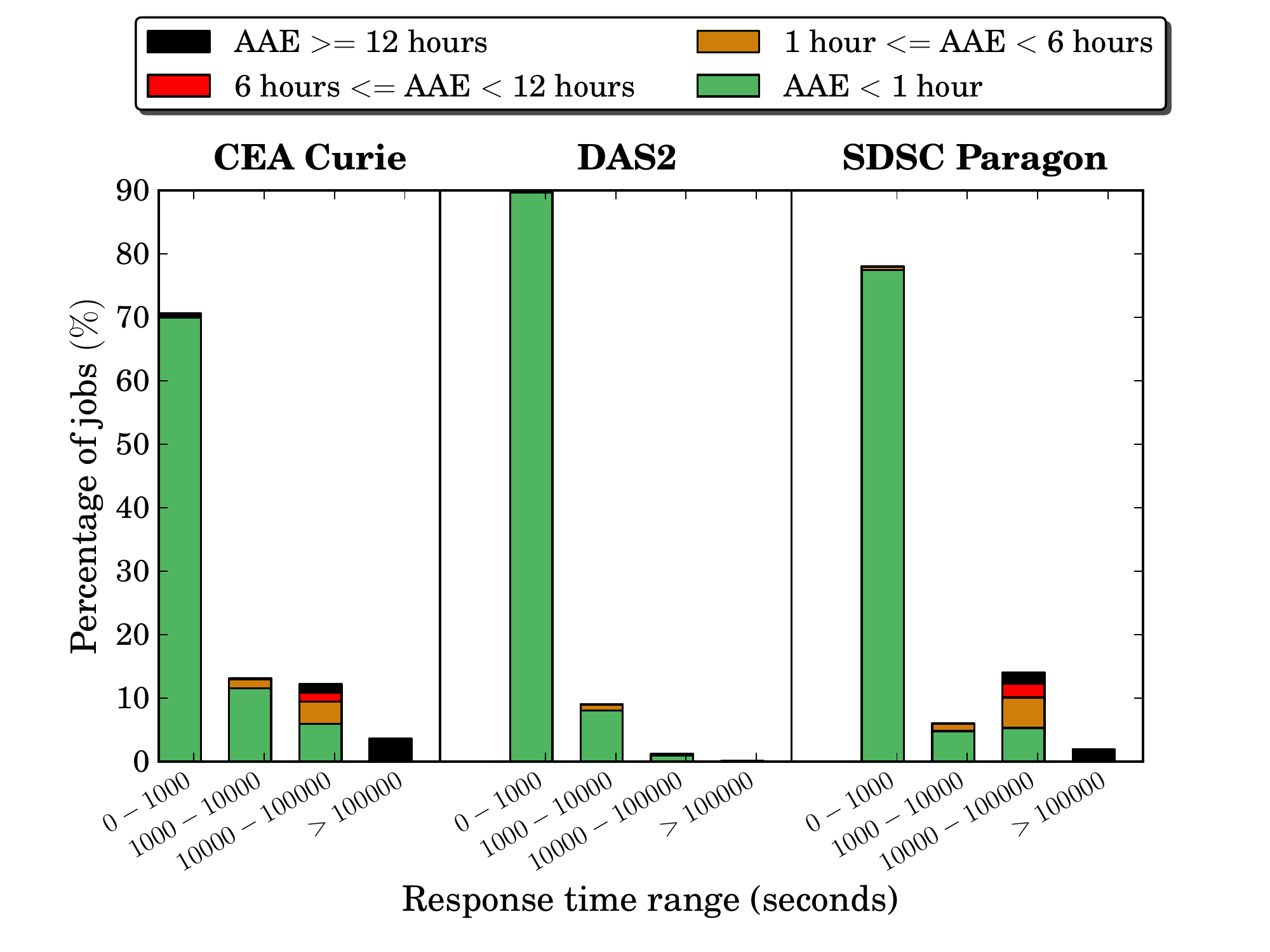}
\caption{\small Distribution of Average Absolute Errors (AAEs) for Different Ranges of Actual Response Times in CEA Curie, DAS2 and SDSC Paragon}
\label{aae_onehour}
\end{figure}

For our current work related to metascheduling, we demonstrate the relevance of our predictor with the given prediction errors for our metascheduling system that uses one-hour day-ahead electricity markets. Figure \ref{aae_onehour} shows the distribution of the average absolute errors (AAEs) for different ranges of actual response times of the jobs for three sample supercomputer traces, namely, CEA Curie of France which is a Top500 system, DAS2 of Netherlands, and SDSC Paragon of USA. The parameters of these supercomputer traces are given in Table \ref{supercomputing_traces}. We find that the AAEs were less than one hour for 88-98\% of the jobs, thus demonstrating that our queue waiting time predictor is sufficiently accurate for metascheduling in day-ahead electricity markets in which prices fluctuate at a frequency of one hour.

\begin{table}
\scriptsize
\centering
\caption{\small Supercomputer Traces}
    \begin{tabular}{|p{0.8in}|p{1.0in}|p{1.2in}|}
    \hline
    Trace name & Trace duration (months) & Number of Completed Jobs \\ \hline \hline
    CEA Curie    & 20 & 266099 \\ \hline
    DAS2         & 12 & 39915 \\ \hline
    SDSC Paragon & 12 & 32199 \\ \hline
    \end{tabular}
 \label{supercomputing_traces}
\end{table}

\subsubsection{Electricity Price Prediction}

\begin{figure}
\centering
 \includegraphics[scale=0.25]{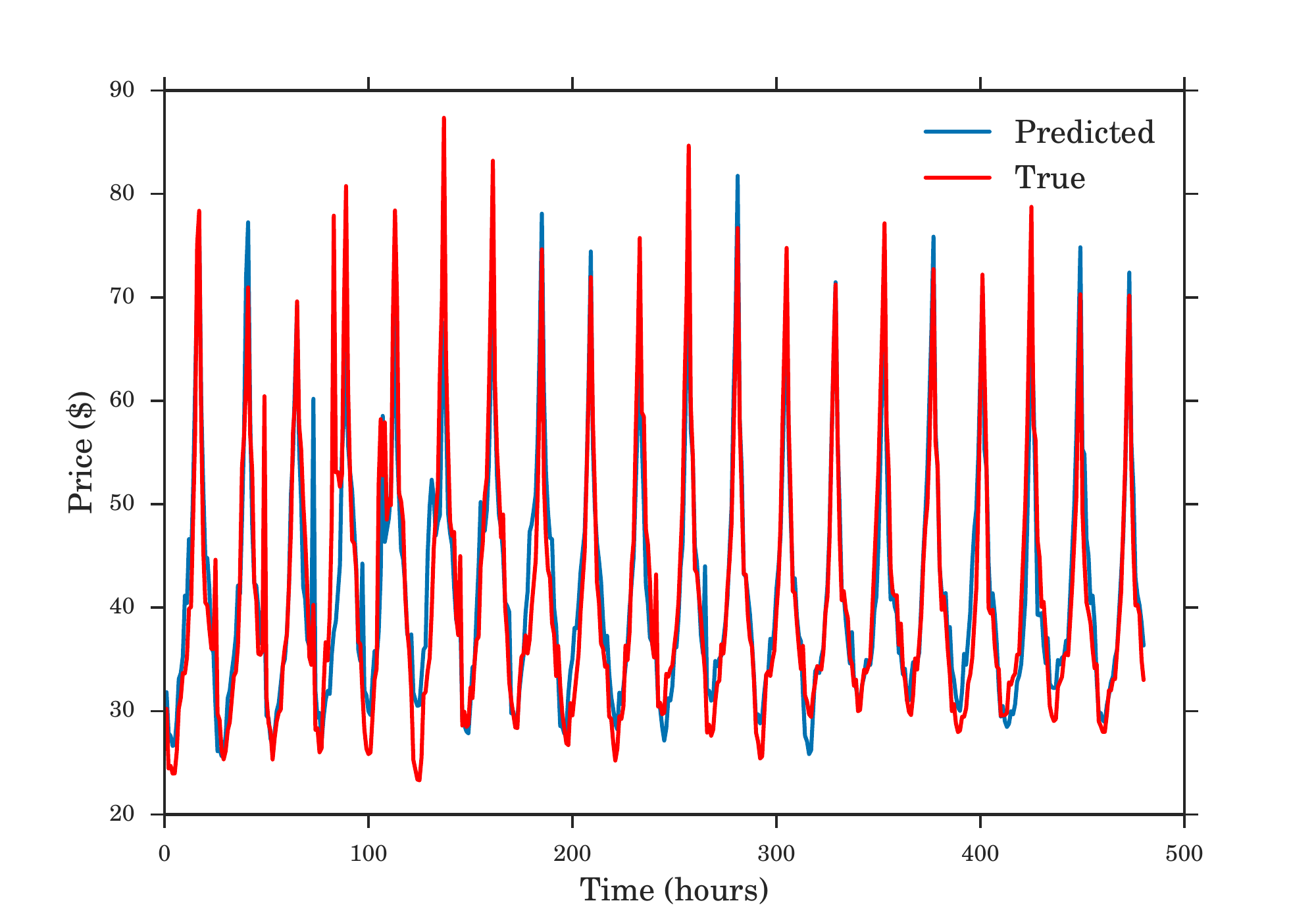}
 \caption{Electricity price predictions for Texas}
 \label{texaspred}
\end{figure}

In Figure \ref{texaspred}, we show the sample prediction results for forecasting of electricity prices in Texas.
For this experiment, we predicted the electricity prices for a single day using the historical prices of the previous three days. The market prices for the day-ahead hourly electricity market were obtained for June 1-20, 2014 from the datasets of Electric Reliability Council of Texas \cite{ercot}. We can see that curves for the predicted and actual prices are very close. The average percentage prediction error was found to be only 8\%.

\subsection{Overall Results}

In this section, we analyze the overall reductions in response times and electricity cost by our algorithm and compare with the other approaches.
Table \ref{Overall-Results-Table} shows the comparison results. The table shows the results of our MCMF algorithm with different values of $w_t$. Recall that $w_t$ denotes the weight of the response time term in the cost function minimized by MCMF.

\begin{table}[htb]
\scriptsize
\centering
\caption {\small Overall Simulation Results}
\begin{tabular}{@{}p{0.4in}p{1.3in}p{0.6in}p{0.6in}@{}}
\toprule
Grid   & Strategy & Average response time (minutes) & Total electricity cost (\$ or \euro) \\ 
\midrule\multirow{5}{*}{XSEDE}
& MCMF $(w_t=25\%)$                & 477.5  & \$224021.6 \\
& \textsc{TWOPRICE}  $(w_t=25\%)$  & 473.3  & \$232557.3 \\
& MCMF $(w_t=0\%)$                 & 1095.4 & \$187298.9 \\			   
& \textsc{INST}                    & 3460.8 & \$205876.5 \\
& Baseline                         & 633.7  & \$230985.6 \\
\midrule
\multirow{4}{*}{NorduGrid}
& MCMF $(w_t=92.5\%)$              & 1678.6 & \euro12819.6 \\
& \textsc{TWOPRICE} $(w_t=92.5\%)$ & 1724.4 & \euro12991.7 \\
& \textsc{INST}                    & 5210.2 & \euro14613.6 \\
& Baseline                         & 1900.3 & \euro16608.3 \\ 
\bottomrule
\label{Overall-Results-Table}
\end{tabular}
\end{table}

\begin{figure}
\centering
  \includegraphics[scale=0.4]{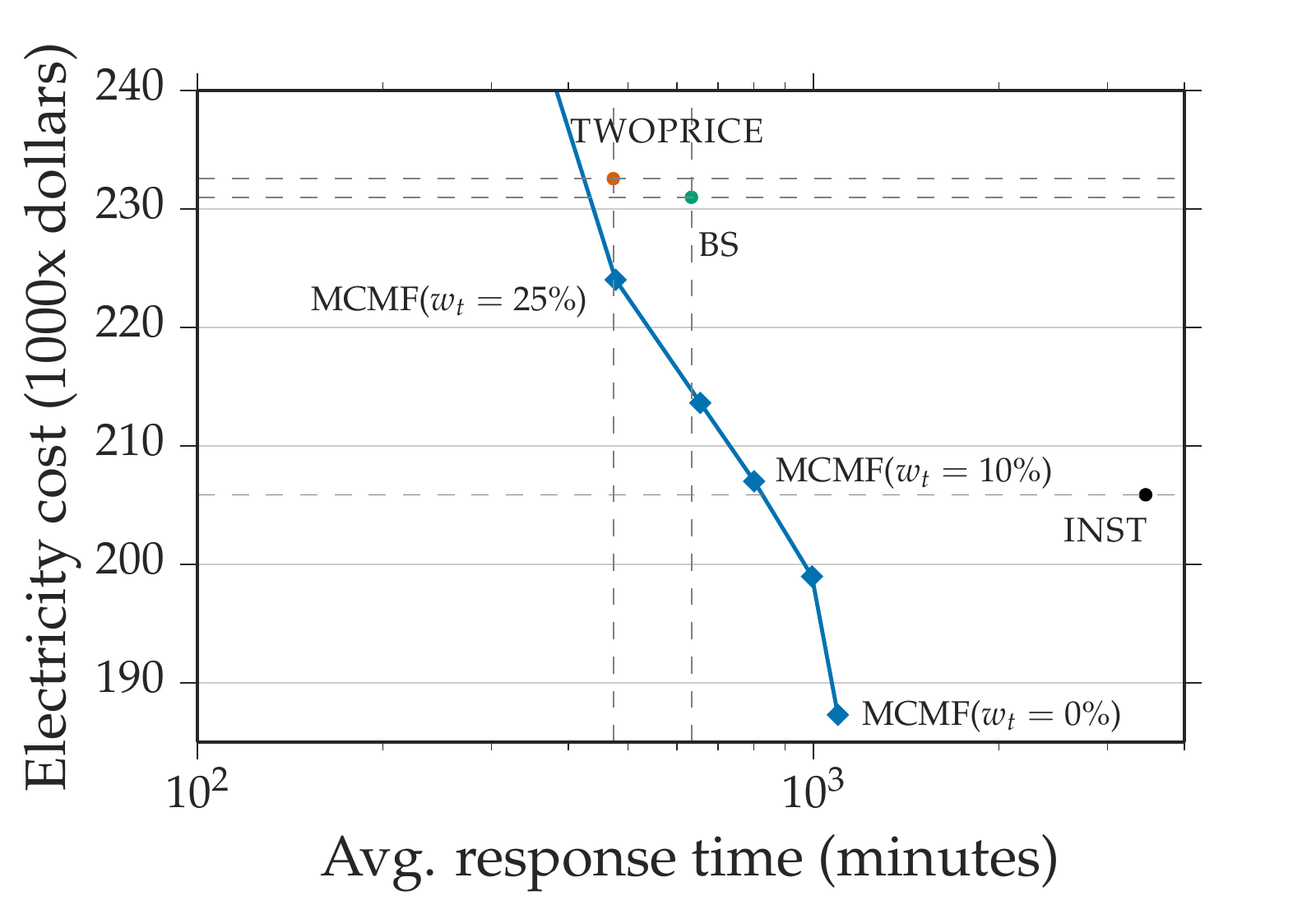}
\caption{Overall simulation results in XSEDE}
\label{XSEDE-OverallResultsFigure}
\end{figure}

For XSEDE, we observe that with $w_t = 25\%$, our MCMF strategy simultaneously achieves 24.6\% reduction in average response time and \$6964 savings in total electricity cost, compared to the baseline, for the 15-day period.  This reduction in electricity cost can potentially translate to a projected savings of \$167K dollars per year for the whole grid. Figure \ref{XSEDE-OverallResultsFigure} shows the trade off between response time and cost for our MCMF algorithm for different values of $w_t$ compared to the other strategies. We see that when response time is not considered for optimization ($w_t=0$), we obtain up to \$43686 reduction in electricity cost with a 1.7x increase in average response time over the baseline. Thus, for one year of operation, we can potentially save \$1.04M for the whole grid. Considering that the annual electricity budget of Argonne National Laboratory's primary supercomputer is \$1M  \cite{Yang:2013:IDP:2503210.2503264}, the savings obtained by our approach are significant.

From Table \ref{Overall-Results-Table} and Figure \ref{XSEDE-OverallResultsFigure}, we can also see that our MCMF algorithm ($w_t=25\%$) outperforms TWOPRICE by \$8535.7 in terms of cost. This is because TWOPRICE is unaware of fine grained price fluctuations every hour.  INST degrades the baseline response time by 5.5x although it achieves better cost.  The reason for INST achieving lower cost and high response times in most of the cases is because at the beginning of the simulation, INST migrates jobs to good systems with low electricity cost and low response times. But soon enough, when the systems become loaded, INST continues to keep pushing jobs to the same systems without being aware of the queue waiting times caused by the high loads on the systems. So being aware of electricity cost helps INST to achieve low cost, while not being aware of waiting time results in high load imbalance across systems, and hence high response times.  Compared to INST, MCMF $(w_t = 10\%)$ obtains 3.1x reduction in response time for the same cost. We also see that our MCMF algorithm ($w_t = 0\%$) outperforms INST in both response time and electricity cost.

Even though INST considers the hourly electricity cost, for jobs with non-zero waiting times it is unable to estimate the electricity prices during the time of job execution without the assistance of a wait time predictor.

For NorduGrid, we observed improvements in both response time and electricity cost when  $w_t = 92.5\%$. For this workload, the response time improves by $11.7\%$ and electricity cost reduced by \euro3788.7 over the baseline. Thus, in NorduGrid, our projected electricity cost savings are \euro15.1K per year. Similar to XSEDE, our MCMF algorithm has lesser response time and electricity cost than TWOPRICE and INST. 

These results show that our MCMF algorithm can achieve the twin goals of reducing both response time and total electricity cost of large scale grids. The results also underscore the importance of both queue waiting time predictions and hourly electricity prices in our MCMF strategy.

\subsection{Sensitivity to Prediction Errors}
\label{sensitivity_predictions}

In this section, we study the effect of prediction errors on the metascheduler. We show results with XSEDE for the 10,000 jobs.

Our electricity price predictions are fairly accurate. This is validated in our experiments by comparing with the actual hourly electricity price data in the day-ahead market for the eight states that constituted the XSEDE grid. In 75\% of the 10,000 jobs, our predictions gave less than 15\% prediction errors. In about all the cases, the predictions gave less than 20\% errors. Since the hourly electricity prices did not vary drastically from one day to the next, our SARIMA model was able to model the prices with reasonable accuracy.

However, the predictions in queue waiting times and runtimes can have large prediction errors for some jobs. As mentioned earlier, the user-estimated runtimes (ERT) we use are generally known to have large prediction errors. Hence in this section, we study the sensitivity of our metascheduler due to the prediction errors in queue waiting and runtime predictions.  For studying sensitivity to prediction errors in queue waiting times, we perform perturbation experiments. For each set of perturbation experiments, we perturb our predicted waiting time for each job by adding a random value in the range $[1,P]$ time units to the initial predicted waiting time, where $P$ is the perturbation threshold.  We perform five sets of perturbation experiments corresponding to thresholds of 1, 3, 6, 12 and 24 hours. We consider the the same set of $10,000$ jobs for each perturbation experiment. Table \ref{sensitivity_qwait_perturb} shows the metascheduling results for jobs for the different perturbation experiments. The first row of the table for the perturbation threshold of 0 hours corresponds to unperturbed results. The table also shows the average PPE in queue waiting time predictions for each perturbation experiment. As expected, the average PPEs increase with increasing perturbation threshold, implying larger errors in queue waiting time predictions for larger thresholds.

\begin{table}
\scriptsize
\centering
\begin{tabular}{|p{0.5in}|p{0.5in}|p{0.8in}|p{0.8in}|}
\hline\hline
Perturbation Threshold (hours) & Average PPE (\%) & Average Response Time (minutes) & Average Electricity Price (\$) \\
\hline\hline
0 & 3 & 458.5 & 22.85 \\ \hline
1 & 25 & 457.9 & 22.90 \\ \hline
3 & 62 & 457.4 & 23.05 \\ \hline
6 & 116 & 462.3 & 23.04 \\ \hline
12 & 221 & 493.6 & 23.02 \\ \hline
24 & 386  & 518.5 & 22.98 \\
\hline\hline
\end{tabular}
\caption{Metascheduling Results for Different Perturbations to Qwait Time Predictions}
\label{sensitivity_qwait_perturb}
\end{table}

We find that the average response times due to our MCMF strategy are relatively stable across different prediction errors, especially when the prediction errors are reasonable. Only for very large percentage prediction errors with average PPEs of greater than 200\% corresponding to thresholds of 12 and 24 hours, we see a noticeable increase in the average response times. As shown in our previous work \cite{prakash-qespera-cpe2016}, the average PPE in our queue waiting time predictions is less than 100\% for most of the real supercomputing traces. We find that the increase in prediction errors did not have an impact at all in the average electricity price yielded by our metascheduler. Thus, our MCMF metascheduler is fairly robust to the prediction errors in queue waiting times.

The user-estimated runtimes already had large prediction errors. Hence, in our original unperturbed experiments, we categorized the 10,000 jobs into different sets corresponding to different ranges of percentage prediction errors in runtimes. For each set of jobs, we then compared the average response times due to our MCMF metascheduler with the other methods. Table \ref{sensitivity_ert} shows the percent improvement or degradation in average response times due to our MCMF metascheduler over the other methods. We find that the improvements or degradations over a particular method does not vary by large amounts with the prediction errors in runtimes. Our MCMF strategy resulted in about 22-35 \% improvement over the baseline for all the sets of jobs. Similarly, the MCMF improvement over INST is in the range 84-90\%, and both the MCMF and the TWOPRICE strategies perform equivalent for all the sets of jobs corresponding to different prediction errors in runtimes. Thus, our metascheduler is also robust to the prediction errors in runtimes.

\begin{table}
\scriptsize
\centering
\begin{tabular}{|p{0.5in}|p{0.5in}|p{0.5in}|p{0.5in}|p{0.5in}|}
\hline\hline
PPE in ERT (\%) & Number of Jobs & \% Imp. over Baseline (\%) &  \% Imp. over INST (\%) &  \% Imp. over TWOPRICE (\%) \\
\hline\hline
0-10 & 1637 & 21.83 & 83.67 & -0.10 \\
\hline
10-20 & 1456 & 25.57 & 88.23 & -0.07 \\
\hline
20-30 & 799 & 25.62 & 88.47 & -0.28 \\
\hline
30-50 & 1070 & 28.81 & 87.36 & -0.18 \\
\hline
50-100 & 1342 & 27.63 & 85.48 & -0.24 \\
\hline
100-200 & 1247 & 29.82 & 87.69 & -0.46 \\
\hline
$>$ 200 & 2448 & 34.56 & 90.18 & -0.35 \\
\hline\hline
\end{tabular}
\caption{Metascheduling Results for Different Ranges of Runtime Prediction Errors}
\label{sensitivity_ert}
\end{table}

\subsection{Effect of Job Size}

\begin{figure}
\centering
\includegraphics[scale=0.4]{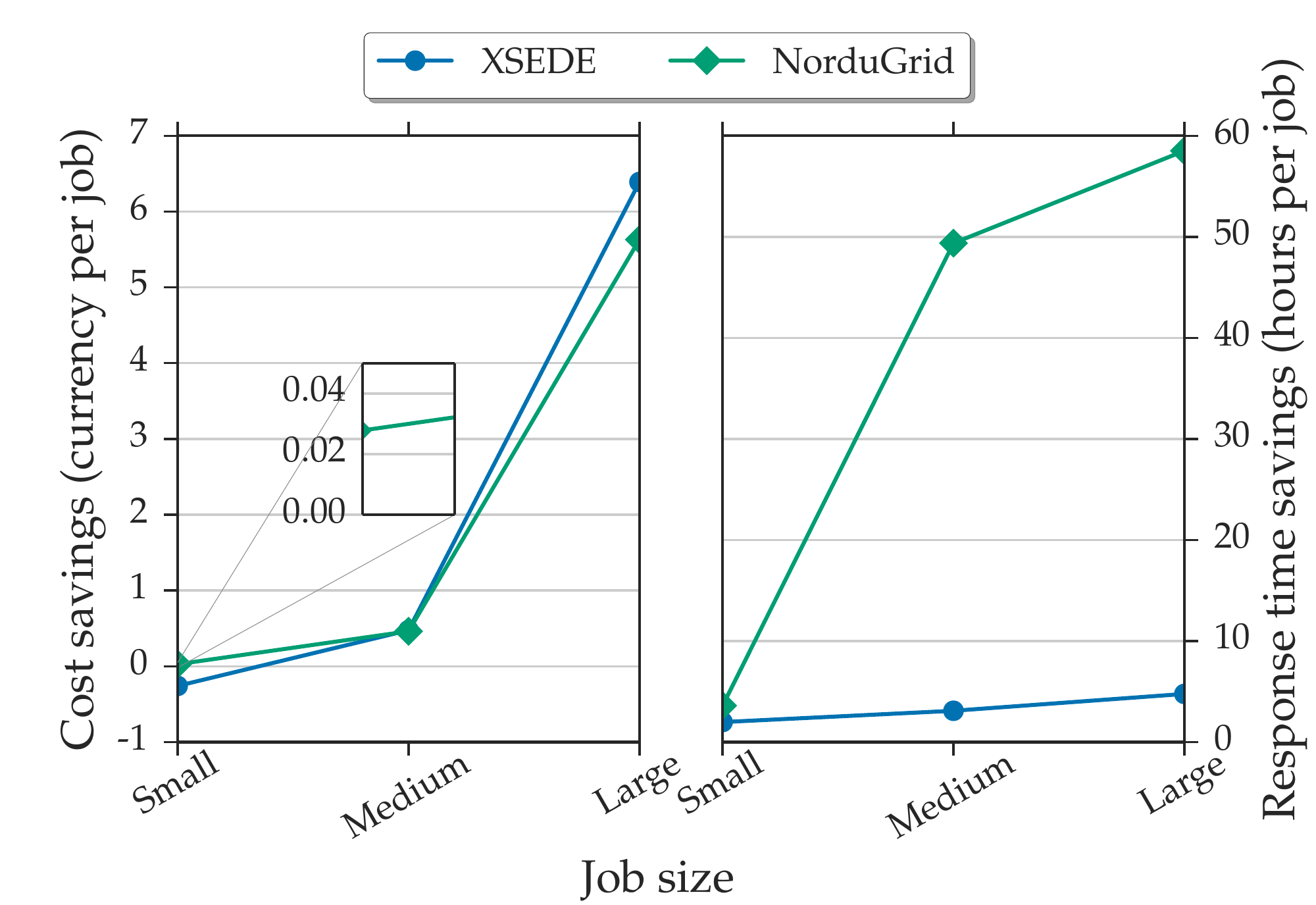}
\caption{Effect on job size on savings}
\label{XSEDE-Savings}
\end{figure}

We also measured how differences in job size impact the savings obtained by our \textsc{MCMF} algorithm over the baseline.  For measuring the impact of job size, we divided the jobs into three classes. We denote jobs having less than 512 CPU hours work as \emph{small}, between 512 and 4096 CPU hours as \emph{medium} and jobs larger than 4096 CPU hours as \emph{large}. In Figure \ref{XSEDE-Savings}, we see that the savings in response time and electricity cost per job increase with job size. Since larger jobs consume more electricity and system core hours, placing these jobs in optimal locations results in larger improvements compared to jobs of smaller size.  As observed in earlier sections, in XSEDE, our algorithm obtains savings in response time by migrating jobs away from slower systems. Similarly, in NorduGrid, we observed that our algorithm obtains high savings in response time as a result of migrating many long running parallel jobs from the Triolith-ATLAS cluster at the National Supercomputer Centre in Sweden which has 2.2GHz Intel Sandy Bridge processors to the Glenn cluster in Chalmers Institute of Technology in Sweden which has 3.0GHz AMD Opteron Interlagos processors.  The difference in the absolute savings in response time for large jobs in XSEDE and NorduGrid arises from the difference in average job runtime in the two grids. On average, jobs in NorduGrid are 3.4x longer than jobs in XSEDE. Hence, migrating the long running jobs in NorduGrid to faster systems gives larger absolute improvements in response time compared to XSEDE.  In XSEDE, we obtain reduction in electricity cost from improved placement of large and medium sized jobs.  It is interesting to note that although NorduGrid exhibits the same trend as XSEDE, 92\% of the total electricity bill saved in NorduGrid is from the optimal placement of small jobs which contribute to the savings of 2.8 cents per job, as shown in the inset in the cost savings graph of Figure \ref{XSEDE-Savings}.  Clearly, even a few cents saved per job can aggregate over the course of months to provide significant reduction in the grid operational cost.

\subsection{Load, Utilization and Power Variations}

In this section, we show a snapshot of the simulations which contrasts the behavior of our algorithm and other approaches.

 \begin{figure}
   \centering
  \includegraphics[scale=0.45]{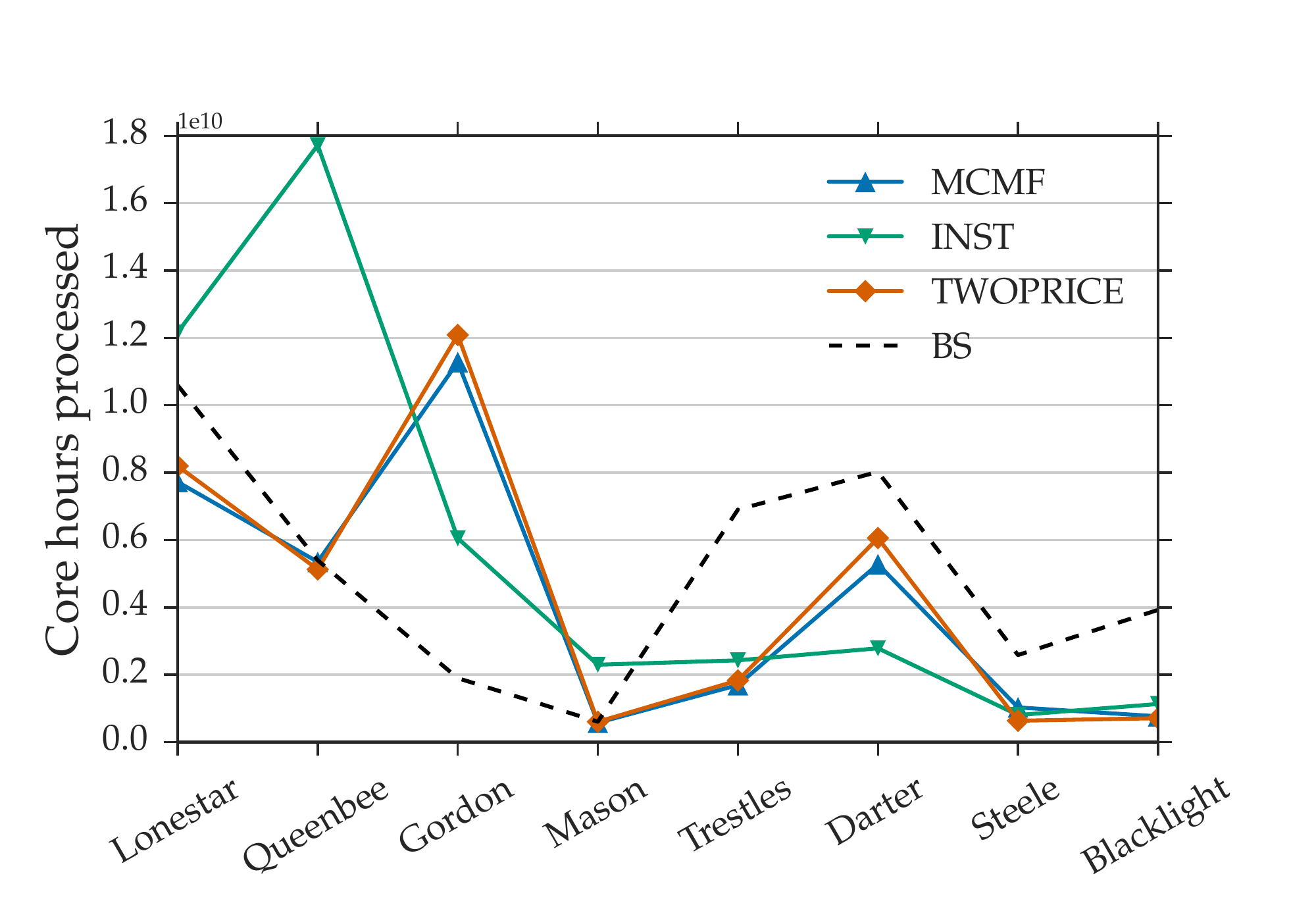}
   \caption{Distribution of delivered core hours among systems in XSEDE}
   \label{Core-Hour-Distribution-W25}
 \end{figure}

In Figure \ref{Core-Hour-Distribution-W25}, we arranged the systems in XSEDE in increasing order of service unit cost and show the the core hours processed at each system. We see that all the electricity price-aware strategies reduce the load on the costly systems compared to the baseline. However, our algorithm is able to achieve the optimal proportion of core hours at each system to obtain improved response time.

\begin{figure}
 \includegraphics[scale=0.4]{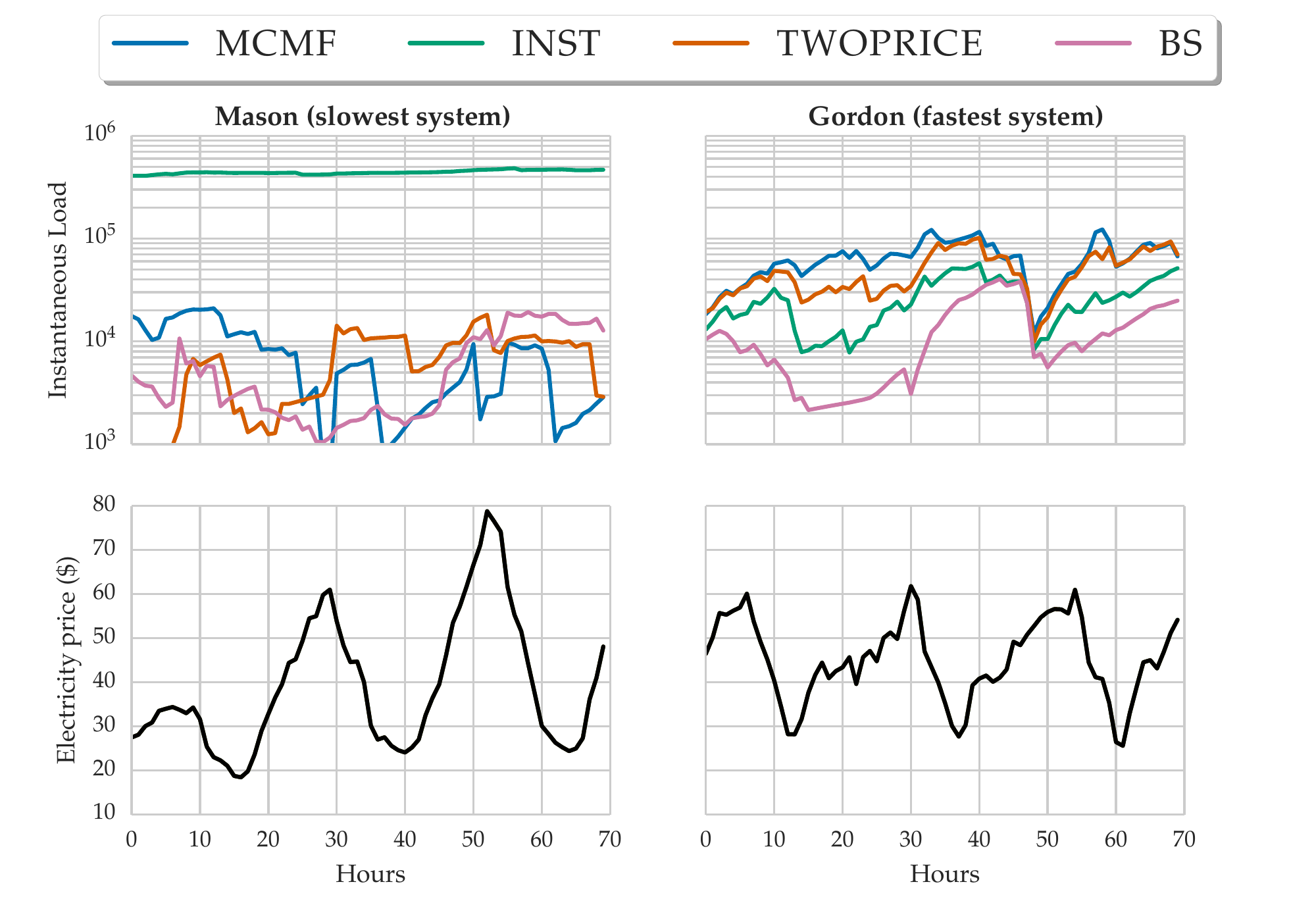}
\caption{Instantaneous Load Variation in XSEDE}
\label{XSEDE-Gordon-Mason}
\end{figure}

We looked at hourly instantaneous load at each system to understand the hourly behavior of our scheduling policy and compared with the other policies. We used $w_t=25\%$ for these experiments.  In Figure \ref{XSEDE-Gordon-Mason}, we contrast the instantaneous load of Mason, the slowest and smallest system in the grid with Gordon, one of the largest and fastest systems. We see that INST achieves very poor load balancing because it is oblivious to response time. We also see that during the peak hours of electricity pricing at Mason, our MCMF algorithm minimizes the instantaneous load among the considered strategies. In Gordon, we see that our approach utilizes the system heavily even during a price peak at hour 30. This is because Gordon has the highest performance and the 3rd lowest service cost among the systems in XSEDE.  We can see that by moving jobs to fast systems which have less service cost, our algorithm is able to simultaneously optimize electricity cost and response time better than the other strategies. 

\begin{figure}
 \includegraphics[scale=0.35]{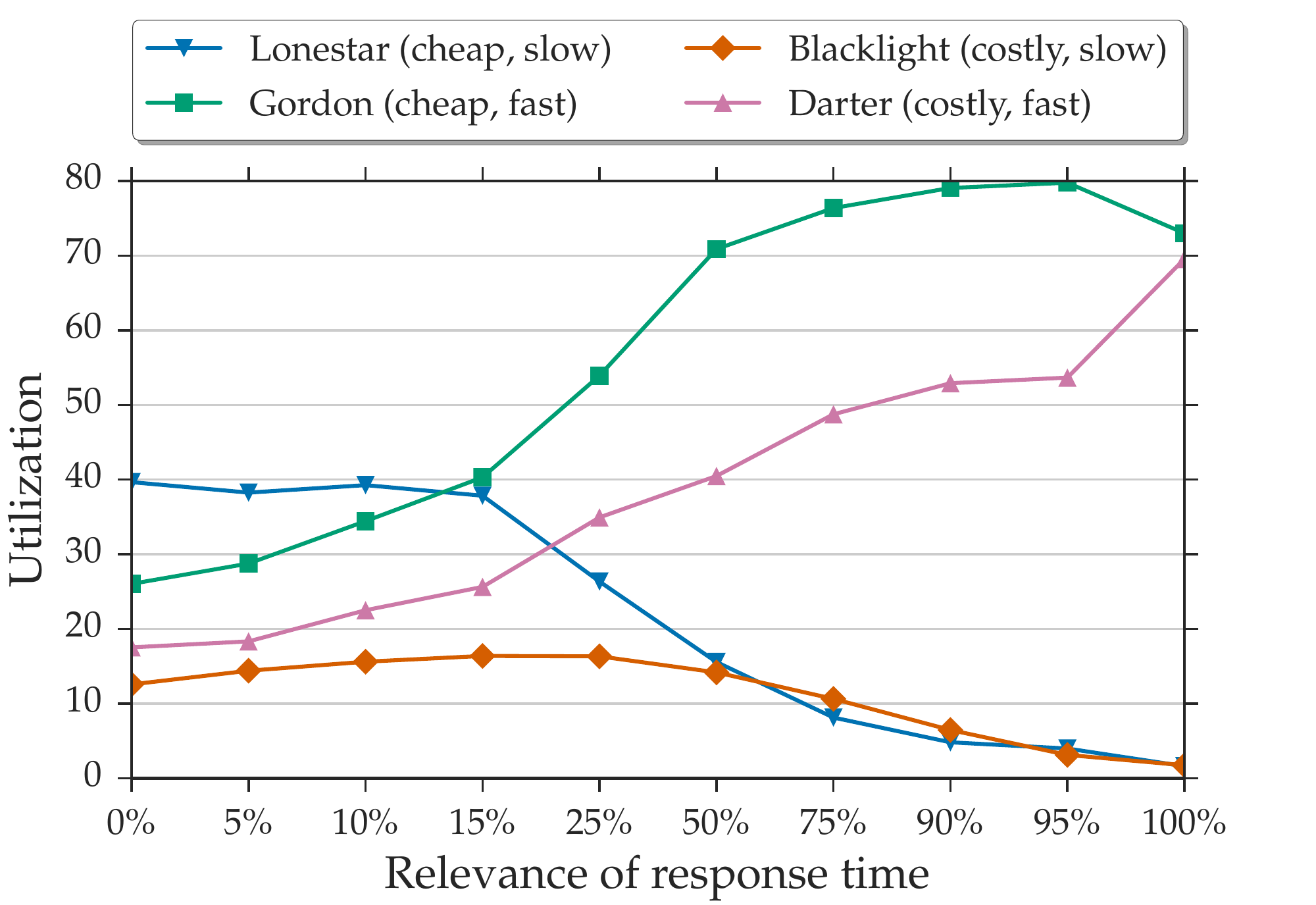}
\caption{Variation of system utilization in XSEDE}
\label{FourSiteUtilization}
\end{figure}

We also studied the variation of overall utilization in different systems due to our metascheduler, and show the results for XSEDE.  We define service unit cost (SUC) as the product of the average electricity price at the system's location and the power per core.  SUC represents the average cost in dollars (or euros) required to deliver one CPU hour of computation at the system.  In XSEDE, we arranged the systems in increasing order of SUC and labelled the first four systems as \textit{cheap} and the remaining as \textit{costly}. Similarly we used the HPL peak performance of the systems to label them as \textit{slow} and \textit{fast}.  In our experiments, we investigated the effect of service unit cost and machine performance on system utilization.  Figure \ref{FourSiteUtilization} shows that as the importance of response time is increased, jobs are migrated from slow systems (Lonestar, Blacklight) to fast systems (Gordon, Darter).  We also see that Gordon has higher utilization than Darter because it has lower SUC. We also note that since Blacklight is relatively costly and slower than the other three systems, it's utilization is low in all the configurations.

\begin{figure}
 \includegraphics[scale=0.4]{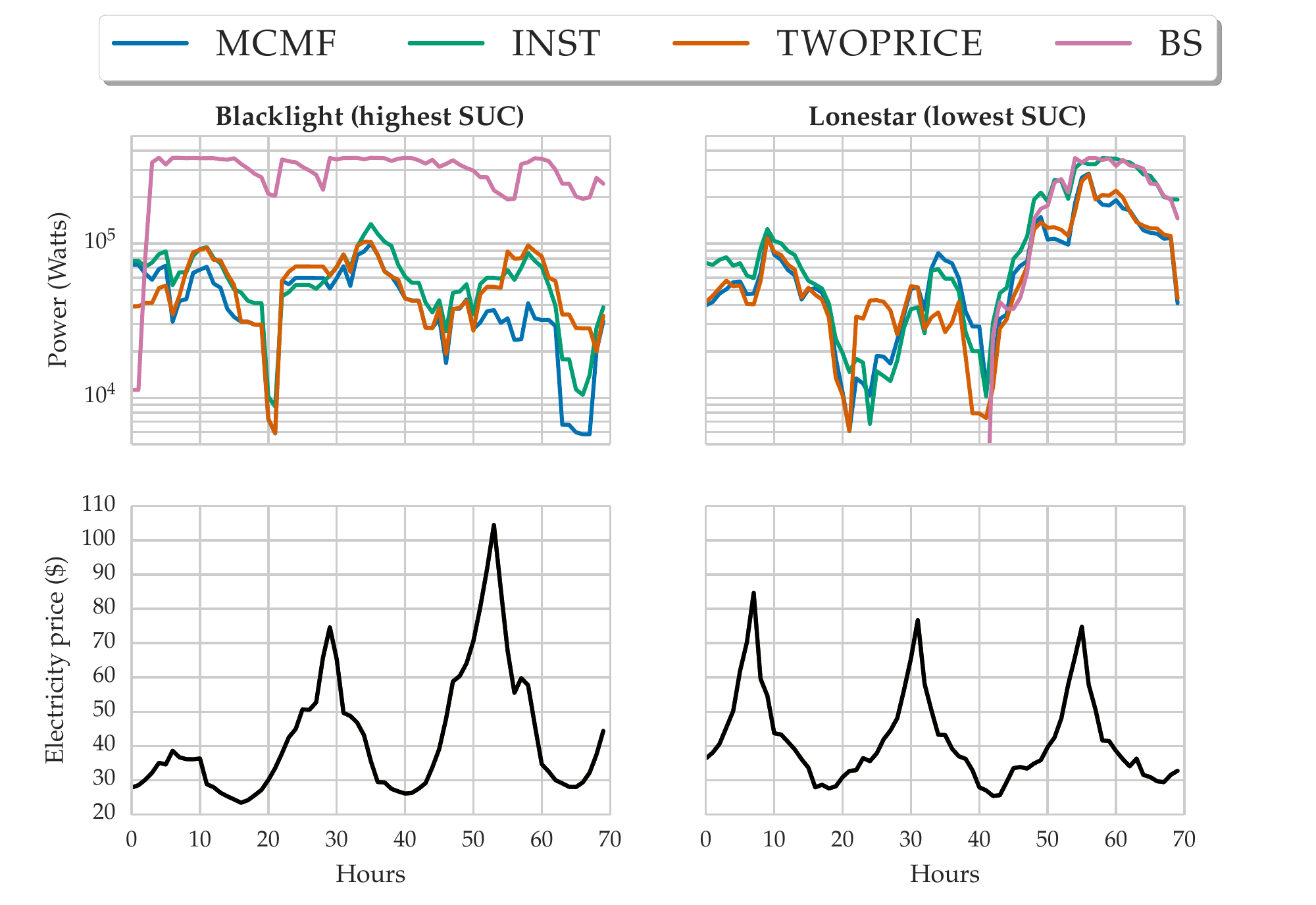}
\caption{Power Consumption Variation in XSEDE}
\label{XSEDE-Blacklight-Lonestar}
\end{figure}

We also observed the hourly power consumption due to the scheduling policies. We used $w_t=25\%$ for these experiments. In Figure \ref{XSEDE-Blacklight-Lonestar}, we compare the hourly power consumption of Blacklight and Lonestar, which are respectively, the costliest and cheapest systems in the grid.  We see that all the electricity price-aware strategies, namely MCMF, INST, and TWOPRICE consume much less power than the baseline in the Blacklight system.  During hours 50-60 when Blacklight experiences peak electricity price, the power consumption of our MCMF algorithm is better than both INST and TWOPRICE.  However, in Lonestar, the fluctuations in electricity price do not influence the load or power consumption significantly even during peak hours of electricity pricing because it is both the largest and the cheapest system in the grid. 

\subsection{Fairness towards individual grid systems}

\begin{figure}
\centering
 \includegraphics[scale=0.25]{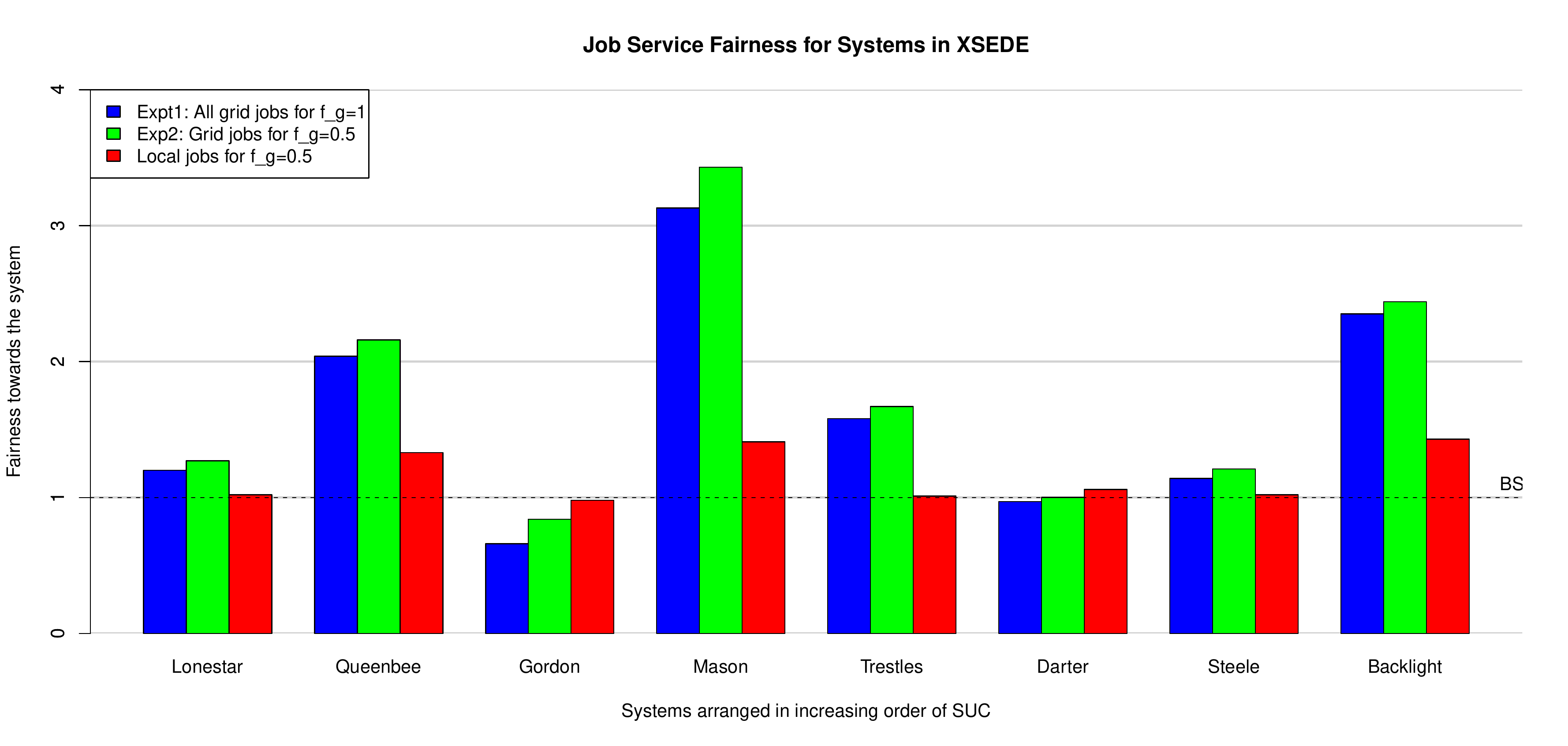}  
\caption{Job service fairness for systems in XSEDE}
\label{xsede-fair}
\end{figure}

In this section, we compute the job service fairness score of each system when user submissions can be either through the metascheduling portal or the local batch system. In one of our experiments for XSEDE, {\em Expt1}, we studied the job service scores when all users submit their jobs through the metascheduler, i.e., every job is a grid submission. The fairness scores for this experiment are indicated in Figure \ref{xsede-fair} by the blue bars. In another experiment, {\em Expt2}, we studied the case where only a subset of the jobs are grid submissions. 

To perform these experiments, we choose a fixed fraction of grid submissions, $f_g$ (e.g., $f_g = 0.5$ denotes that 50\% of the jobs are submitted to the metascheduler), and for each job submission, we conduct a single Bernoulli trial with probability of success equal to $f_g$.  Jobs with successful trials are routed through the metascheduler and the remaining jobs are considered as submissions to local batch system.  We performed the experiments for $f_g = 0.5$, repeated each run 5 times and averaged the scores using geometric mean. In Figure \ref{xsede-fair}, the green bars indicate the fairness scores for the 50\% grid submissions and the red bars indicate the fairness scores for the 50\% local batch queue submissions.  We indicate the service fairness of the baseline strategy with a line which is labelled as BS. Service fairness scores more than 1 indicate improved response times compared to the baseline.

When all jobs are grid submissions, we can see that all systems have values more than 1 except Gordon. This indicates that
jobs which originated at these systems obtained benefits in response time due to metascheduling. Gordon, which has service fairness slightly less than 1, is the fastest and 3rd cheapest system in the grid. In the baseline strategy, the average response time of jobs is close to zero, i.e., no waiting in the queue. Hence, our MCMF algorithm migrates many jobs to this system. But, the jobs processed at Gordon incur an average waiting time of only half an hour, which indicates that the users of this system did not suffer much due to grid participation. Jobs which originated at slow smaller sized systems like Queenbee, Mason and Blacklight, obtained large benefits from metascheduling. 

When only a subset of the jobs are grid submissions, we see that both users of the grid and the local batch system obtained benefits in response time. Grid users obtained improved performance because of job migration. Local users obtained improved performance at systems like Queenbee, Mason and Blacklight because grid submissions were migrated away from these systems, leaving more resources free to process local submissions. Thus, we see that a system's participation in a grid which uses our metascheduling algorithm, provides benefits even for users who do not submit through the grid portal.

\subsection{Sensitivity to Metascheduling Parameters}

We studied the effect of three important parameters of our algorithm: $w_t$, $MaxQ$ and the percentage of grid submissions.  Recall that $w_t$ denotes the weight of the response time term in the cost function minimized by MCMF and $MaxQ$ represents the number of jobs that MCMF can schedule at a system in one scheduling cycle. Varying $w_t$ and $MaxQ$ allows us to study the structure of the optimization space and provides insights which can be used for making scheduling policy decisions.

\subsubsection{Varying $w_t$}
\begin{figure}
\centering
\includegraphics[scale=0.45]{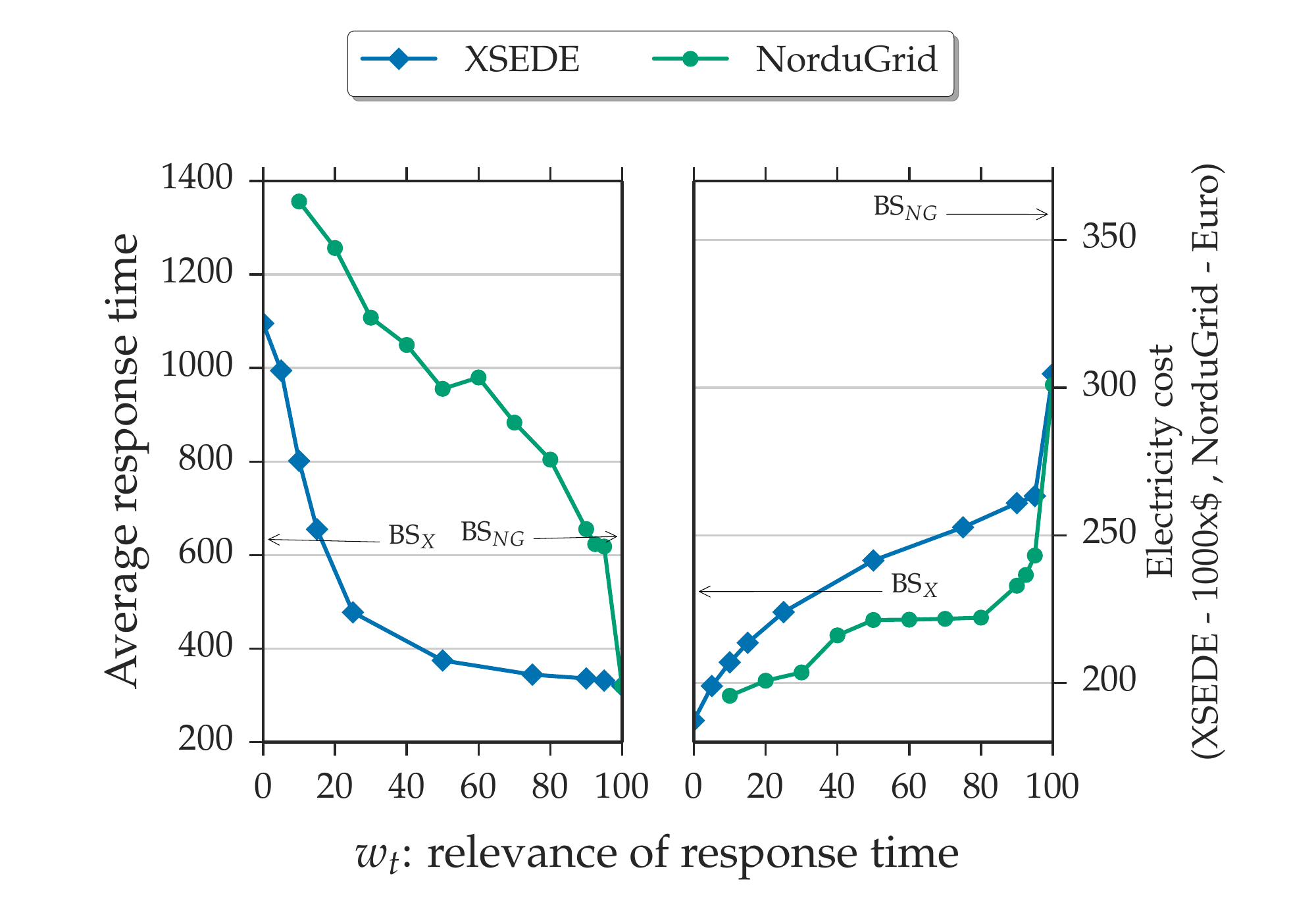}
\caption{Effect of varying $w_t$ on response time and cost}
\label{XSEDE-NorduGrid-WT-Variation}
\end{figure}

Varying $w_t$ allows the grid administrators to control the relative importance of minimizing response time vs minimizing total electricity cost. These objectives can be conflicting in the presence of daily fluctuations of electricity price. If response time is not important, it can be traded off to run more jobs during hours with lesser electricity price. Figure \ref{XSEDE-NorduGrid-WT-Variation} shows the effect of varying $w_t$ in XSEDE and NorduGrid using 10000 jobs.  For response time and electricity cost, $BS_X$ and $BS_{NG}$ represent the baseline value in XSEDE and NorduGrid, respectively.  We see that increasing the relevance of response time (electricity cost) leads to a decrease in response time (electricity cost). We observed that when only response time is minimized ($w_t=100\%$) we are able to obtain $48-49\%$ reduction in response time over the baseline in both XSEDE and NorduGrid.

Similarly, when only electricity cost is considered ($w_t=0\%$), our scheduling strategy obtains $18\%$ and $46\%$ reduction in total electricity cost in XSEDE and NorduGrid, respectively. It is interesting to note that, in NorduGrid, for all values of $w_t$, our MCMF algorithm outperforms the baseline in terms of electricity cost.  In XSEDE, we can see that for $w_t$ values between 20-40\%, both response time and electricity cost are better than the baseline. So, we selected $w_t=25\%$ as the optimal value for XSEDE. In NorduGrid we selected $w_t=92.5\%$. Compared to XSEDE, in NorduGrid, we require a high value of $w_t$ to get improvements in response time. This is because the cost function minimized by MCMF is skewed depending on the magnitude of the response time and electricity cost. In NorduGrid, we observed that average runtime is 3.4x greater than XSEDE and the average electricity cost of a job is 177x lesser than XSEDE. The optimal values of $w_t$ are different in XSEDE and NorduGrid because of the differences in the range of reponse times and electricity costs in each grid.  Grid administrators can use a test workload to obtain these trends using our framework and decide an appropriate value of $w_t$ depending on the budget and user service agreements. 

\subsubsection{Varying the percentage of grid submissions}

Typically, large scale grids expose their resources to users with a local batch scheduler at each system and a global metascheduling system which facilitates remote job submission.  Grid administrators also partition their resources for local and remote submissions to offer differentiated job service classes. In this section, we investigate the effect of limiting the percentage of grid job submissions.

\begin{figure}
 \includegraphics[scale=0.4]{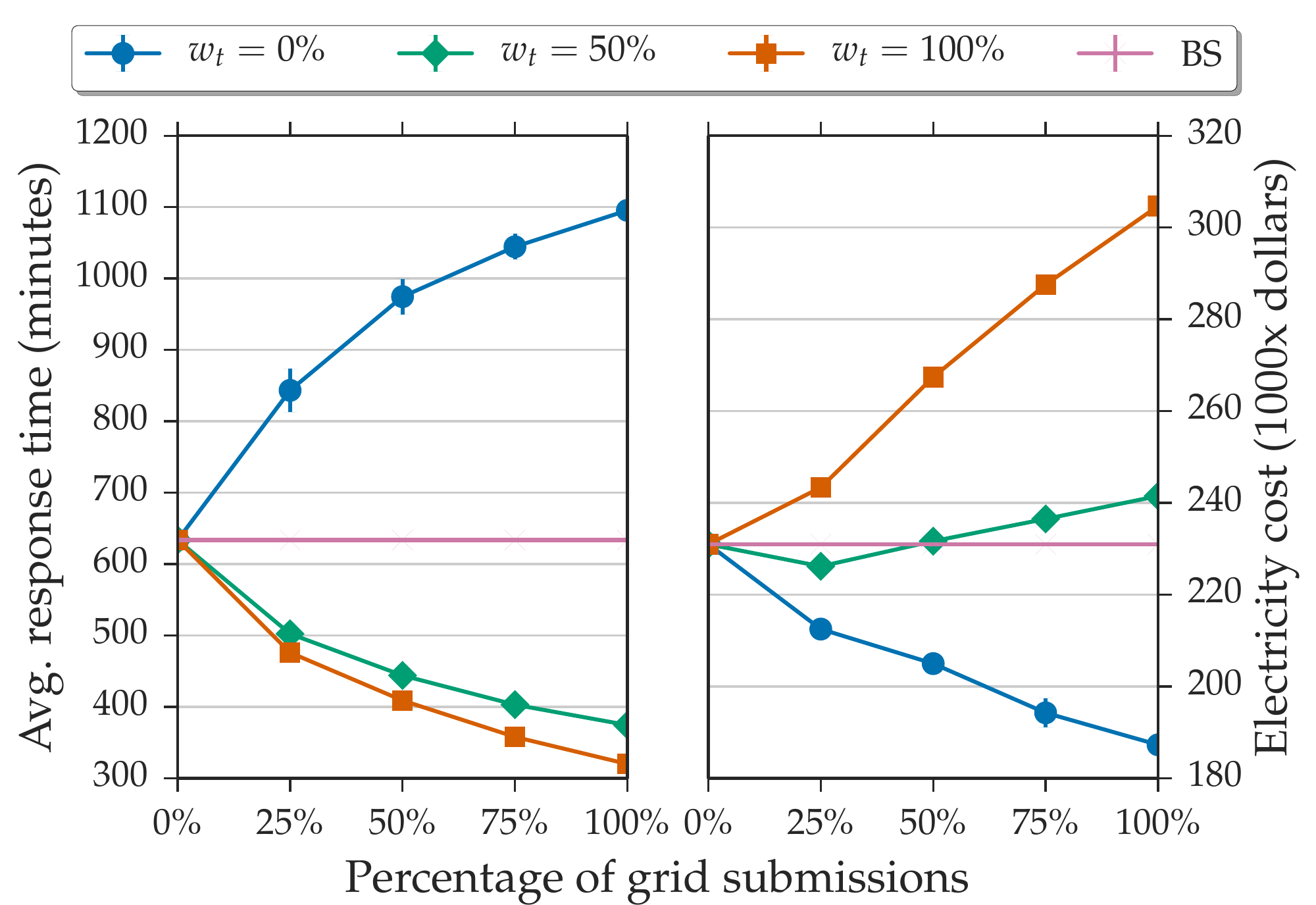}
\caption{Tradeoffs observed for different $\%$ of grid submissions in XSEDE}
 \label{Grid-META-Variation}
\end{figure}

We performed experiments for different fractions of grid submissions, $f_g$. Each result with a given $f_g$ corresponds to an average of five runs.  To perform these experiments, we choose a fixed fraction of grid submissions, $f_g$ (e.g., $f_g = 0.5$ denotes that 50\% of the jobs are submitted to the metascheduler), and for each job submission we conduct a single Bernoulli trial with probability of success equal to $f_g$.  Jobs with successful trials are routed through the metascheduler and the remaining jobs are considered as submissions to local batch systems.  We repeated each run 5 times and performed the experiment for different values of $f_g$.  The results for XSEDE are shown in Figure \ref{Grid-META-Variation}.  In each graph, we indicate the response time/cost of the baseline strategy with a line which is labelled as BS.  Across different tests, we can see that the reduction in response time or electricity cost is proportional to its relevance in the cost function and the percentage of grid submissions.  We see that when the relevance of response time is high, more grid submissions is beneficial because our algorithm has more jobs which can be migrated to better systems. Similarly, when electricity price is important, grid submissions results in lesser total electricity price across the grid.  We see that even with a small percentage of grid submissions we gain benefits in response time and electricity cost compared to the baseline strategy, when the cost function considers response time ($w_t\ne0\%$) and electricity cost ($w_t\ne100\%$), respectively.  Since the error bars at each point in the graph are small, it implies that the improvements are not sensitive to the exact subset of jobs chosen for the experiments.  With 100\% job submissions through a metascheduler and for wt=50\%, we gain 40.8\% reduction in average response time with almost the same electricity cost as the BS. Hence, for further experiments, we use $f_g=1.0$ for both XSEDE and NorduGrid.  

\subsubsection{Varying MaxQ}

\emph{MaxQ} which denotes the number of jobs that can be submitted to a system during a scheduling cycle determines the total number of jobs that the metascheduler can dispatch in a cycle. It is important to choose \emph{MaxQ} carefully because when queue waiting time is predicted for a job, the predictor is not aware of the other jobs that may be submitted to the same system during the same scheduling cycle. Hence, allowing a large value for MaxQ can lead to worsening of response times because of errors in the queue waiting time predictions.

\begin{figure}[h]
\centering
  \includegraphics[scale=0.3]{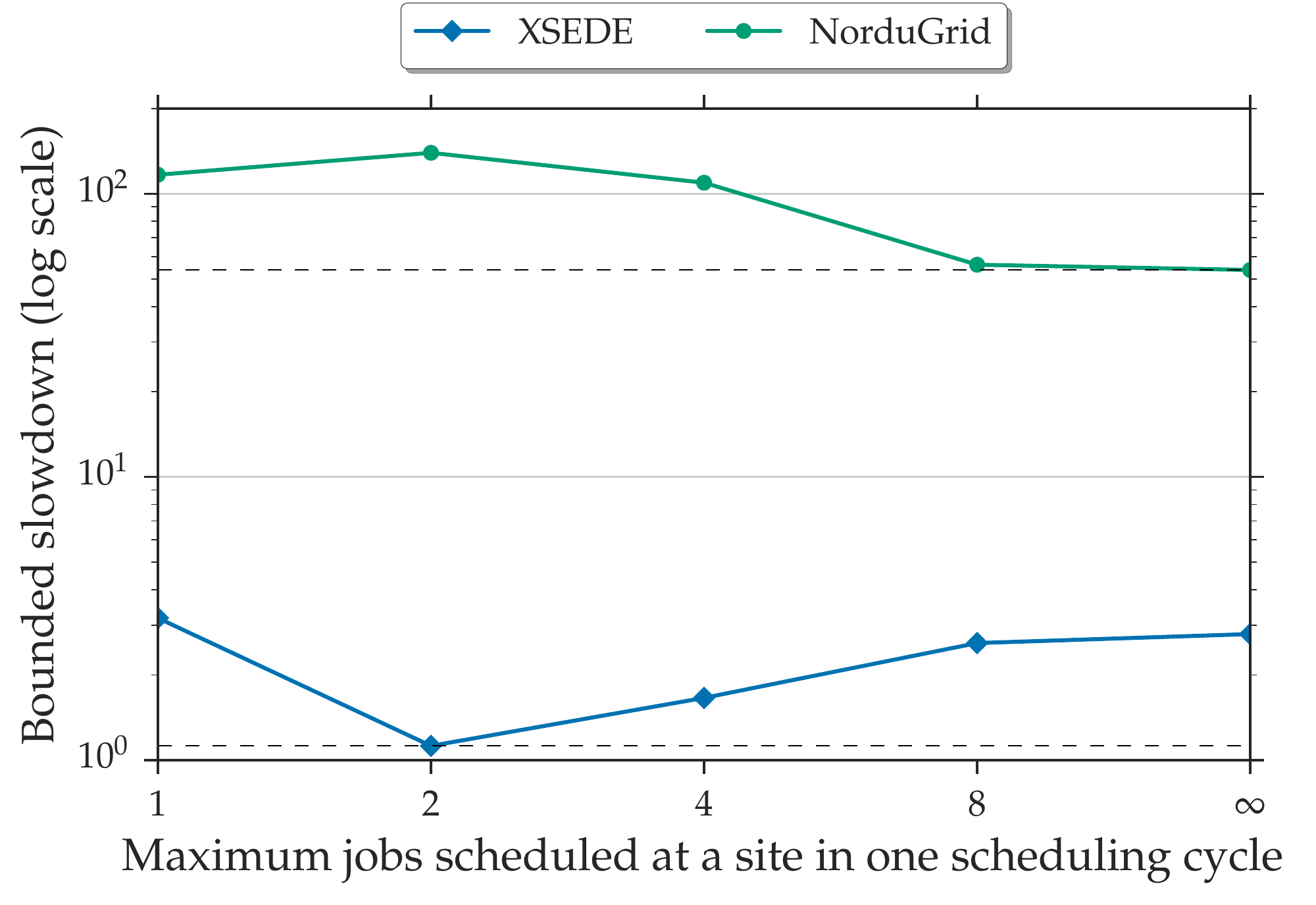}
\caption{Effect of varying \emph{MaxQ}}
  \label{VaryingMaxQ}
\end{figure}

For XSEDE, we observed that average response time reduces when MaxQ is increased from 1 to 2 because jobs are not held in the metascheduler queue. When MaxQ is further increased the average response time increases. We also observed that the trend is more pronounced in the bounded slowdown metric which is shown in Figure \ref{VaryingMaxQ}. We use $MaxQ=\infty$ to denote the case where we do not impose any limit on the number of job submissions to a system in a scheduling cycle.  For the NorduGrid workload, we observed that increasing MaxQ improves the average response time and bounded slowdown even with $MaxQ=\infty$.  This behavior arises from the difference in average inter-arrival time of the two workloads. In XSEDE, the average inter-arrival rate is less than 4 jobs per hour compared to 67 jobs per hour in NorduGrid. So each scheduling cycle in NorduGrid receives significantly more jobs than XSEDE and large MaxQ allows the scheduler to submit more jobs to individual systems in each cycle. Based on these observations, we choose $MaxQ$ as 2 and $\infty$ in XSEDE and NorduGrid respectively.

\subsection{Power Consumption and Data Communication Models}

Our previous experiments did not consider data transfers between the submission and the execution site. In this section, we consider a data transfer and communication model in which data movement from the submission site is initiated simultaneously with the job migration and submission to the execution site. We extended our execution model to include the data transfer time as:
\begin{equation}
responseTime = (max(commT, qwT) + execT)
\end{equation}
where $qwT$ is the queue waiting time on the site to which the job is migrated and executed (execution site), $commT$ is the time for communication of data between the submission to the execution site, and $execT$ is the execution time in the submission site. Our metascheduler used this response time to make its decision.

Our previous results were also obtained with the assumption that the power consumption by the applications is the same as the consumption by HPL. This is based on studies using comprehensive simulation by Kamil et al. \cite{4536223}. Subsequently, real experiments with diverse set of applications on large scale systems in the work by Laros et al. \cite{laros-energyPerfTun-hpc2012} and Song et al. \cite{song-energyprofiling-ijhpca2009} suggest that the power consumption can vary between -40 to +40\% of HPL's power consumption, with no specific skew towards higher or lower values. In this section, we experimented with three different power consumption models: an {\em average-HPL-power} model in which we randomly chose the power consumption of a job in the range of -40\% to +40\% of HPL's power consumption, a {\em lower-than-HPL-power} model corresponding to the range -40\% to 0\% in which 0\% corresponds to using the HPL's power consumption, and a {\em higher-than-HPL-power} model corresponding to the range 0 to +40\%.

We conducted experiments using 10K jobs, and involving both the above mentioned communication and power consumption models. In the first set of experiments, we show the effect of the communication models in our results. For this, we chose the power consumption model as {\em average-HPL-power} model. For each job in our experiment, we randomly chose the data size of the job as one of 0, 1 KB, 1 MB, 10 MB, 100 MB, 500 MB, 1 GB, 10 GB, 100 GB, 500 GB, 1 TB, and 10 TBytes. We used the latency and grid-ftp bandwidth data available in \cite{perfsonar-data} and \cite{psc-gridftp} for the communication links in XSEDE. Table \ref{comm_results_table} shows comparison results with the communication model.

\begin{table}
\scriptsize
\centering
\begin{tabular}{||p{1.1in}|p{1.0in}|p{0.75in}||}
\hline\hline
Strategy & Average response time (minutes) & Total electricity cost (\$) \\
\hline\hline
MCMF $(w_t=25\%)$ &  518.2 & 229784.7 \\
\hline
Baseline & 678.4 & 225085.7 \\
\hline
MCMF $(w_t=20\%)$ & 640.7 & 220479.4 \\
\hline
TWOPRICE $(w_t=25\%)$ & 499.0 & 234341.7 \\
\hline
INST & 1172.5 & 201411.5 \\
\hline\hline
\end{tabular}
\caption {Simulation Results involving Communication Model}
\label{comm_results_table}
\end{table}

Similar to the overall results shown in Table \ref{Overall-Results-Table}, we find similar comparisons when including network transfer times. With $w_t$ set to 25\%, we find that when compared to the baseline MCMF gives reduction of 150 minutes in average response time. However, the electricity cost due to MCMF is about \$4K more than the cost due to the baseline. But the advantage of MCMF is that it can be tuned to suit the needs of a supercomputer site.
By setting its $w_t$ parameter to 20\%, we find that it outperforms the baseline in both the average response time and the electricity cost. 
MCMF also outperforms the TWOPRICE method in terms of electricity cost with savings of more than \$5K with only a 20-minute increase in average response time.
The TWOPRICE algorithm obtains worse electricity cost than MCMF because it does not consider fine grained variations in electricity price.
Similar to the earlier results of Table \ref{Overall-Results-Table}, MCMF gives large-scale reductions in response times when compared to INST while giving higher electricity cost. The INST algorithm which does not consider queue waiting time suffers from large response times. However its performance is better than the earlier case of Table \ref{Overall-Results-Table} since considering network bandwidth allowed it to move jobs away from systems with low network bandwidth.

We now show the effect of different power consumption models on the results. For these experiments, we restricted the data size to 1 TBytes. Table \ref{power-variations-table} shows the comparisons. We find that with the variations in the power consumptions across the rows, the response times show only small-scale or even negligible variations in all the methods. The total electricity costs, as expected, increase across the rows almost uniformly for all the methods. Thus, MCMF continues to maintain its relative position wrt the other methods irrespective of the power consumption model: its average response time is about 2 hours less than that of the baseline and less than half of the response time of the INST method, and its electricity cost is about \$5K less than the cost of the TWOPRICE method for comparable response times.

\begin{table*}
\scriptsize
\centering
\begin{tabular}{||p{1.0in}|p{0.5in}|p{0.5in}|p{0.5in}|p{0.5in}|p{0.5in}|p{0.5in}|p{0.5in}|p{0.5in}||}
\hline\hline
Power Model & \multicolumn{2}{|c|}{Baseline} & \multicolumn{2}{|c|}{INST} & \multicolumn{2}{|c|}{TWOPRICE} & \multicolumn{2}{|c||}{MCMF} \\
\hline
& Avg. resp. time (mins.) & Elec. Cost (\$) & Avg. resp. time (mins.) & Elec. Cost (\$) & Avg. resp. time (mins.) & Elec. Cost (\$) & Avg. resp. time (mins.) & Elec. Cost (\$) \\
\hline\hline
{\em lower-than-HPL-power}  & 638.99 & 220766.14    & 1171.25 & 197803.79     & 462.65 & 230634.33    & 467.82 & 225568.33 \\
\hline
{\em average-HPL-power}     & 638.99 & 225085.79    & 1110.72 & 199142.72     & 458.82 & 233869.51    & 465.51 & 228040.67 \\
\hline
{\em higher-than-HPL-power} & 638.99 & 229355.91    & 1153.75 & 201035.58     & 456.47 & 236758.28    & 466.05 & 231575.37 \\
\hline\hline
\end{tabular}
\caption {Results for Different Power Models}
\label{power-variations-table}
\end{table*}

\subsection{Practical Considerations}

In each scheduling cycle, the meta scheduler collects information about the queue and processor status of each system in the grid and the current list of pending jobs. This information is processed by our MCMF algorithm and the jobs are submitted to the appropriate systems.  From the web statistics published by NorduGrid \cite{nordu-access}, we observed the information From the web statistics published by NorduGrid, we observed the information collection phase takes less than 30 seconds for all the 80 systems in the grid. During our experiments, we observed that our Python implementation running on an Intel Core i7 3.4Ghz processor with 16GB RAM takes $8.4$ seconds on average for computing the scheduling cost and constructing the flow network, and $16.3$ seconds on average for computing the minimum cost flow and the subsequent job submissions to individual systems. Assuming that a scheduling cycle happens every few minutes \cite{xsede-condor}, we conclude that our implementation is fast enough to be deployed in currently operational grids.

\section{Related Work}
\label{related-work}

 In this section, we present different classes of related work and describe why previous efforts cannot address the present problem.  Approaches which reduce power consumption by lowering CPU frequency or voltage \cite{5493460} may not be widely and uniformly applicable across the entire grid due to the autonomous systems that are involved. Hence we do not describe related works which primarily employ such techniques to achieve power savings.

\textit{Single HPC system scheduling:}

The works of Yang et. al.\cite{Yang:2013:IDP:2503210.2503264} and Zhou et. al.\cite{JSSPP-power-aware} formulate the electricity price aware job scheduling problem for a single computing system as a 0-1 knapsack model.  These works do not use hourly electricity prices. Instead, they consider two electricity price values corresponding to on and off-peak hours.  Their algorithm is applied during peak hours to maximize utilization while maintaining the power consumption within a power budget that is specified a-priori.  Our work is different from these approaches because we consider a connected grid of computing systems and route job submissions through a metascheduler. We also consider hourly electricity pricing and have shown improvements over a strategy which uses only on-peak and off-peak prices.

\textit{Datacenter scheduling:}

The concept of {\em geographic load balancing} \cite{Liu:2011:GGL:1993744.1993767} has been used for distributing Internet traffic across distributed data centers. Qureshi et.al. \cite{Qureshi:2009:CEB:1592568.1592584} proposed electricity price aware request routing for Akamai's web traffic workload. Liu et. al. \cite{Liu:2011:GGL:1993744.1993767} proposed geographic load balancing of Hotmail traffic requests to achieve energy savings.  Rao et al.\cite{5461933} use minimum cost flow for scheduling service requests in geographically distributed Internet data centers.  Ren and He developed COCA \cite{export:199682}, a scheduling framework which uses Lyapunov optimization to minimize operational cost of the data center while satisfying carbon neutrality constraints.  Each server which is modelled as a M/G/1/PS (Memoryless/General/1/Processor-Sharing) queue adjusts the amount of workload it can process and the processing speed to dispatch jobs at a particular service rate. This work uses one hour ahead electricity price prediction.

These approaches are applicable only for Internet data center workloads and not batch system workloads. They assume that requests are uniform with similar service times and employ techniques which use overall request arrival and service rate statistics.  In a typical HPC or grid workload, requests/jobs are highly non-uniform in terms of running time, requested number of processors and queueing delay.  These works consider that the request is serviced in the submission hour and do not consider requests which require many hours or days of computation. Thus, the combination of workload and service policy used in HPC centers cannot be accurately modeled by these previous works.  Our work predicts the execution period of a job using a history based queue waiting time predictor and considers actual/predicted electricity prices during this future period.  These previous works also do not capture the queue dynamics at HPC/grid systems which use space sharing scheduling algorithms with policies like backfilling and fairshare scheduling unlike the FIFO request queues in Internet datacenters.

\textit{Grid scheduling:}

Rathore and Chana \cite{gridsurvey} provide a detailed survey on the state of the art in load balancing and job migration techniques used in grid systems.  England and Weissman \cite{england-costandbenefits-jsspp2004} have studied the benefits of sharing parallel jobs in computational grids for both homogeneous and heterogeneous grids.  Our work considers a more realistic model of the grid sites governed by batch queueing policies. In addition, we also consider electricity prices.  Mutz and Wolski \cite{MutzW08}, developed auction based algorithms for implementing job reservations in grid systems.  Chard et. al. \cite{4534260} proposed an auction based scheduling framework where participating virtual organizations collaboratively arrive at scheduling decisions.  Sabin et. al. \cite{sabin1} proposed a metascheduling algorithm based on the multiple simultaneous reservations at different systems in a heterogeneous multi-site environment.  Subramani et. al. \cite{1029936} proposed greedy load balancing strategies which schedule jobs > at a subset of the least loaded sites.  None of these previous works are cognizant of electricity price or job power characteristics. To our knowledge, ours is the first work on metascheduling HPC workloads across grid systems which optimizes both response time and electricity cost.

\section{Conclusions}
\label{conclusion}

The operational cost of large scale computing grids is expected to increase with the demand for more computational power and time across various scientific domains. In this paper, we presented a Minimum Cost Maximum Flow based formulation of the grid scheduling problem to optimize the total electricity price and average response time of HPC jobs in large scale grids operating in day-ahead electricity markets. Using two currently operational computational grids,  

\begin{small}
\bibliographystyle{IEEEtran}
\bibliography{pricing}
\end{small}

\begin{IEEEbiography}[{\includegraphics[width=1in,height=1.25in,clip,keepaspectratio]{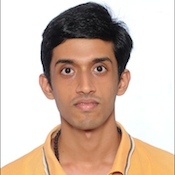}}]{Prakash Murali} is currently doing PhD in the computer science department at Princeton University, USA. He received a Masters degree from Indian Institute of Science, Bangalore (IISc). In IISc, he was advised by Prof. Sathish Vadhiyar and he worked on problems related to scheduling in HPC systems. He subsequently worked as a software engineer at IBM Research, India in the areas of parallel graph and tensor algorithms and workload scheduling in datacenters.
\end{IEEEbiography}

\begin{IEEEbiography}[{\includegraphics[width=1in,height=1.25in,clip,keepaspectratio]{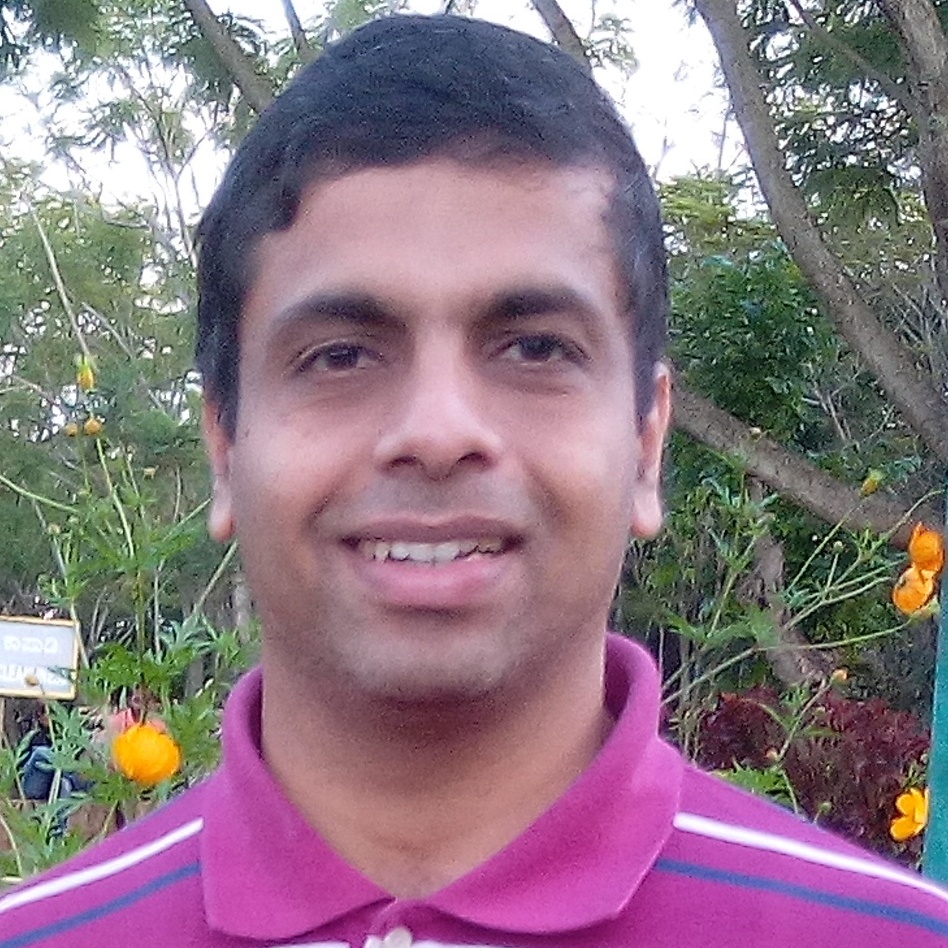}}]{Sathish Vadhiyar}
is an Associate Professor in Supercomputer Education and Research Centre, Indian Institute of Science. He obtained his B.E. degree from the Department of Computer Science and Engineering at Thiagarajar College of Engineering, India in 1997 and received his Master's degree in Computer Science at Clemson University, USA in 1999. He graduated with a Ph.D from the Computer Science Department at University of Tennessee, USA in 2003. His research areas are building application frameworks including runtime frameworks for irregular applications, hybrid execution strategies, and programming models for accelerator-based systems, processor allocation, mapping and remapping strategies for Torus networks for different application classes including irregular, multi-physics, climate and weather applications, middleware for production supercomputer systems and fault tolerance for large-scale systems.
\end{IEEEbiography}

\end{document}